\documentclass[twocolumn]{aastex631}

\usepackage{natbib}
\usepackage[version=4]{mhchem}
\usepackage{enumitem}
\usepackage{threeparttable}

\definecolor{background}{RGB}{252, 229, 241}
\definecolor{frame}{RGB}{225, 34, 138}

\newlist{gitemize}{itemize}{4}
\setlist[gitemize,1]{
  leftmargin=\dimexpr1.5cm+\labelsep\relax,
  label={\smash{\raisebox{-0.25\height}{\includegraphics[width=0.6cm]{planet2_trans.png}}}}
}

\newcommand{\RNum}[1]{\uppercase\expandafter{\romannumeral #1\relax}}

\DeclareSymbolFont{UPM}{U}{eur}{m}{n}
\DeclareMathSymbol{\umu}{0}{UPM}{"16}
\let\oldumu=\umu
\renewcommand\umu{\ifmmode\oldumu\else\math{\oldumu}\fi}
\newcommand\micro{\umu}

\let\microns \jmicron

\shorttitle{Determining the Detectability of \ce{H2O} with Photometry}
\shortauthors{Latouf et al.}

\begin{document}
\title{Determining the Detectability of \ce{H2O} with Photometric Observations using Bayesian Analysis for Remote Biosignature Identification on exoEarths (BARBIE)}

\author[0000-0001-8079-1882]{Natasha Latouf}
\affiliation{NASA Postdoctoral Program Fellow, NASA Goddard Space Flight Center, 8800 Greenbelt Road, Greenbelt, MD 20771, USA}
\affiliation{Sellers Exoplanents Environment Collaboration, 8800 Greenbelt Road, Greenbelt, MD 20771, USA}

\author{Chris Stark}
\affiliation{NASA Goddard Space Flight Center, 8800 Greenbelt Road, Greenbelt, MD 20771, USA}
\affiliation{Sellers Exoplanents Environment Collaboration, 8800 Greenbelt Road, Greenbelt, MD 20771, USA}

\author[0000-0002-8119-3355]{Avi M. Mandell}
\affiliation{NASA Goddard Space Flight Center, 8800 Greenbelt Road, Greenbelt, MD 20771, USA}
\affiliation{Sellers Exoplanents Environment Collaboration, 8800 Greenbelt Road, Greenbelt, MD 20771, USA}

\author[0000-0002-5060-1993]{Vincent Kofman}
\affiliation{Centre for Planetary Habitability (PHAB), Department of Geoscience, University of Oslo, Norway}

\correspondingauthor{Natasha Latouf}
\email{nlatouf@gmu.edu, natasha.m.latouf@nasa.gov}

\begin{abstract}

We examine the detectability of water (\ce{H2O}) in the reflected-light spectrum of an Earth-like exoplanet assuming a photometric observational approach rather than spectroscopic. By quantifying the detectability as a function of normalized exposure time, resolving power (R), and amount of spectral points, we can constrain whether spectroscopy or photometry is the more efficient observing procedure to detect \ce{H2O} at varying abundances by measuring the broad 0.94 {\microns} absorption feature using the Habitable Worlds Observatory (HWO). We simulate low-resolution spectroscopy (R = 10, 20, 30, presented as photometric bandwidth fraction 10\%, 5\%, 3\% herein) as a proxy for narrow-band photometric observations, and constrain the wavelength range from 0.85 - 1.05 {\microns}, to narrow focus on the 0.9 {\microns} feature. We then constrain the number of spectral points to 2 or 3 points at each bandwidth fraction to investigate the impact of spectral point placement on detectability. Additionally, we take the signal-to-noise ratios (SNRs) for strong \ce{H2O} detection and calculate the resultant exoplanet yields assuming photometric observation and compare to the yields from higher-resolution spectroscopic observations under different noise instances, characterization wavelength, noise floors, and aperture sizes. We find that \ce{H2O} is strongly detectable at all bandwidth fractions depending on the spectral point placement, requiring a minimum of 3 spectral points, at a variety of normalized exposure time depending on the abundance of \ce{H2O}. We also find that the detector noise is the main driver in determining whether photometry or spectroscopy results in higher yields. Photometry is the preferred observational method in high-noise cases, while spectroscopy is preferred in low-noise scenarios.

\end{abstract}


\section{Introduction}
\label{sec:intro} 

Since the first planet was found orbiting a Sun-like star by \citet{mayor95}, over 5,000 exoplanets have been discovered and confirmed. The new age of exoplanet research is taking shape, with the detection and characterization of terrestrial exoplanet atmospheres underway with JWST \citep{alderson23, beatty24, espinoza25} and planned with future observatories. Detecting and confirming \ce{H2O} in an exoplanet atmosphere is typically thought of as the first step in the characterization of potentially Earth-like exoplanets \citep{luvoir}, and terrestrial habitable zone (HZ) planets for which \ce{H2O} is detected will be priority re-observation targets. 

When considering the design of next-generation instruments, mission concept studies place \ce{H2O} detection at the forefront of key science goals for the characterization of Earth-like planets. Proposed future generations of observatories, including LUVOIR \citep{luvoir} and HabEx \citep{habex}, adopted \ce{H2O} detection as a part of the initial search for exoEarth candidates (EECs), followed by the detection of \ce{O2}, \ce{O3}, \ce{CH4}, and many other biosignatures and contextual molecular signatures. LUVOIR and HabEx have now evolved into the Habitable Worlds Observatory (HWO) mission \citep{decadal}, and \ce{H2O} detection remains a critical component of the observation strategy. Throughout the development of LUVOIR, HabEx, and the upcoming HWO, the assumed method of observation for biosignature detection has always been spectroscopy, using either an Integral Field Spectrograph (IFS) or fiber-fed spectrographs. The required spectral resolving power, R, has typically been assumed to be 140 at visible wavelengths\citep{luvoir}, driven primarily by work by \citet{brandt14}. \citet{brandt14} estimated the resolving power necessary to detect both \ce{H2O} and \ce{O2}, arriving at $R=150$ for \ce{O2}. As a result, it has been assumed that the \ce{O2} band would drive the design of the IFS at visible wavelengths, resulting in an adopted spectral resolution of $\sim$150. 

However, \citet{brandt14} also noted that \ce{H2O} could be detected at a lower resolving power, approximately R $\simeq$ 40 at 0.94 {\microns} and even lower at longer wavelengths; subsequent work by \citet{robinson20} found a very similar resolving power for the detection of \ce{H2O}, of $\sim$45 at 0.94 {\microns}. However, the FWHM of the \ce{H2O} line at 0.94 {\microns} is 0.074 {\microns} (or 74 nm), which may not require IFS observations. Photometrically detecting \ce{H2O} with multi-wavelength photometry could be advantageous, as an IFS can reduce the end-to-end throughput and increase the number of detector noise pixels. Detecting \ce{H2O} using photometry could therefore result in shorter exposure times. Understanding which molecules, particularly \ce{H2O}, are accessible with photometry will be critical in ensuring efficient observing procedures. Although photometric observations are taken individually, by retrieving on multiple points simultaneously, the retrieval can perform more accurately with context and break the inherent degeneracies in single-filter observations \citep{ducrot24}.


In order to understand \ce{H2O} as a function of resolving power, we use the spectral grid-based Bayesian inference method first described in \citet[hereafter S23][]{susemiehl23}, and used in \citet[hereafter BARBIE1][]{latouf23}, \citet[hereafter BARBIE2][]{barbie2} and \citet[hereafter BARBIE3][]{barbie3}. We then run a set of yield calculations to understand the effect of photometric observation of \ce{H2O} on the resultant exoEarth yields. In this work, we constrain the detectability of \ce{H2O} as a function of resolving power to better understand if photometry is a more efficient method to detect \ce{H2O}. In $[\S]$ \ref{sec:method} we present the methodology of our simulations, through both the BARBIE retrievals and yield calculations. In $[\S]$ \ref{sec:results} we present the results of our retrievals of \ce{H2O} and the corresponding yield calculations. In $[\S]$ \ref{sec:disc} we discuss the results and analyze the possibility of utilizing photometry as an alternative to IFS observations to detect \ce{H2O}. In $[\S]$ \ref{sec:conc} we present our conclusions.

\section{Methodology}
\label{sec:method}

\begin{figure*}[ht!]
\centering
\includegraphics[scale=0.5]{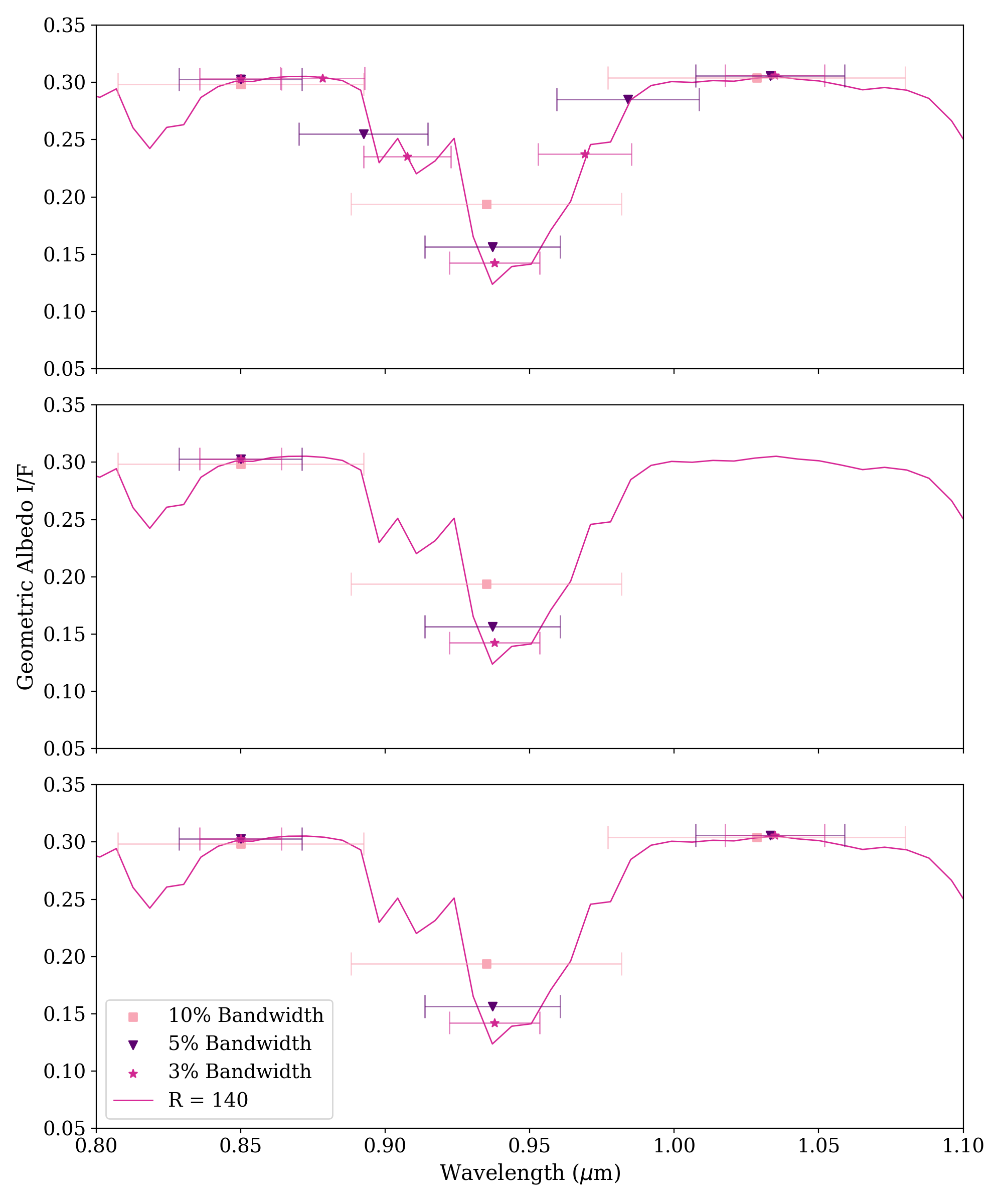}
\caption{The \ce{H2O} absorption feature with our simulated photometric points overlaid. Plotted are a spectrum at R = 140 resolving power, spectral points at 10\%, 5\%, and 3\% bandwidths in light pink squares, purple triangles, and dark pink stars respectively. The width of the wavelength bin is shown in horizontal lines for the respective photometric points in the bottom panel. \textbf{\emph{Top panel:} our full spectra with all sequential points.} \emph{Middle panel:} our two-point photometry observation scenario. \emph{Bottom panel:} our three-point photometry observation scenario. All consist of a 50\% cloudy model. The x-axis is wavelength, constrained from 0.8 to 1.05 {\microns}, and the y-axis is geometric albedo (I/F). \textbf{At wider bandwidths, the feature is more smoothed.}}
\label{fig:fullspec_rcomp}
\end{figure*}


For this study, we use the Planetary Spectrum Generator \citep[PSG,][]{PSG,PSGbook} to generate all simulated spectra. PSG is a radiative transfer model for creating the simulated planetary spectra, spanning a wide range of wavelengths (50 nm to 100 mm) and incorporating a full suite of atmospheric constituents and radiative processes (including includes aerosol, atomic, continuum (i.e.~CIAs, Rayleigh, Raman), and molecular scattering and radiative processes). The HITRAN molecular absorption parameters \citep{gordon20} are used as the base molecular opacities. PSG was also used to produce the spectral grids used in BARBIE. We use PSG to generate spectra at a variety of resolving powers ranging from 50 to 140, and also to simulate individual photometric data points. For simulated spectra, we use binned spectra and ignore the instrumental line spread function. For the \ce{H2O} feature, this approximation is valid because our range of resolving powers exceeds the intrinsic \ce{H2O} line width. For simulated photometric data points, we convert the desired bandwidth to a spectral resolving power (i.e. a 10\% bandwidth is an R of 10, following that bandwidth is equal to 1/R) and pull out individual points from simulated PSG binned spectra, effectively assuming a top-hat filter profile with unity transmission


 
Prior studies have shown that the 0.94 {\microns} \ce{H2O} feature is likely the most efficient absorption feature to detect in the atmosphere of a potentially Earth-like planet \citep{stark24a}. We therefore limit our wavelength range from 0.85 to 1.0 {\microns} for this study. In prior BARBIE studies, we varied the molecular abundances and signal-to-noise ratios (SNRs), and investigated R=70 and 140. Our current study varies not only the molecular abundances, but also varies the photometric bandwidth through 10\%, 5\%, and 3\% bandwidths. We consider two photometric observing scenarios, illustrated in Figure \ref{fig:fullspec_rcomp}: a two-point scenario that has one point at the depth of the feature and one point in the continuum short-ward of the feature, and a three-point scenario that adds a photometric point in the continuum long-ward of the feature. We additionally vary the SNR from 3 to 20 to understand the SNR necessary for detection and draw conclusions on the efficacy of photometric observations. 

In order to directly compare the results for varying bandwidths and observational methods to each other, we present all results in normalized exposure time instead of SNR as in prior BARBIE works, following Equation~\ref{eq:tnorm}:

\begin{equation}
\label{eq:tnorm}
   T_{norm} = \frac{R_1}{R_2} \cdot (\frac{SNR_{1}}{SNR_{2}})^{2}
\end{equation}

Where R$_2$ = 30, SNR$_{2}$ are the SNRs associated with R = 30 (i.e. integer values between 3 -- 20), and R$_1$ SNR$_{1}$ are the R and associated SNRs, respectively. In this way, all results are unified to a common metric for presentation. Equation~\ref{eq:tnorm} makes the assumption that the dominant noise contribution is photon noise (assumed in the $\frac{SNR_{1}}{SNR_{2}})^{2}$ term), thus following Poisson statistics, and scales as the square of the number of photons, i.e. the square of exposure time. 

Figure~\ref{fig:fullspec_rcomp} demonstrates the photometric spectral points used as compared to a spectroscopic, R = 140 spectrum. The solid line shows a 50\% cloudy spectrum at R = 140, which is the currently accepted resolving power in the VIS wavelength regime, and the resultant spectral points from a 10\% bandwidth (light pink squares), to a 5\% bandwidth (purple triangles), and a 3\% bandwidth (dark pink stars). We can see that by binning the spectrum to extreme low resolving powers, fewer spectral points are produced and each point has a larger bandwidth. We can also see the intrinsic biases that are introduced at wider bandpasses, such as the higher albedo at the \ce{H2O} feature depth at 10\%, that approaches the true feature depth as the bandwidth decreases. 

Figure~\ref{fig:fullspec_rcomp} portrays the placement of points when all points are included (top panel), to showcase the points that are removed in our tests. The middle panel shows the spectra when constrained to only 2 points, one at continuum shorter than the \ce{H2O} feature and on at the depth of the feature. For a 5\% and 3\% bandwidth, the spectral points are non-sequential (i.e. points were removed to constrain to two points). In the bottom panel, a point is at continuum on either side of the feature depth. As before, 5\% and 3\% bandwidths have both had points removed in order to limit to 3 points, one at the continuum on either side of the feature, and one at the center of the deepest point in the spectral feature. This adds more realism to our methodological proxy for photometric observations. Additionally, we ensure as much as possible that each point is at the same wavelength, for a true direct comparison (i.e. the only difference is the width of the bin per spectral point). We ran a series of two tests varying the amount of spectral points as seen in Figure\ref{fig:fullspec_rcomp}. In this way, we are able to explore which variable has the greatest impact on \ce{H2O} detectability: spectral point placement, bandwidth fraction, or normalized exposure time. However, in order to account for the photometric serialization penalty in exposure time, Equation~\ref{eq:tnorm} must be adapted slightly. We assume that the 2-point scenario is our baseline, and thus Equation~\ref{eq:tnorm} must be multiplied by $3/2$ to account for the additional penalty of observing a third point. Thus, Equation~\ref{eq:tnorm} becomes

\begin{equation}
\label{eq:tnorm_3pt}
   T_{norm} = \frac{3}{2} \cdot (\frac{R_1}{R_2} \cdot (\frac{SNR_{1}}{SNR_{2}})^{2})
\end{equation}

to calculate the normalized exposure time for the 3-point scenario such that the results can be directly compared to the 2-point scenario.

\subsection{Spectral Retrievals}

We use a similar methodological approach to the spectral retrieval analyses as our prior studies BARBIE1, BARBIE2, and BARBIE3. See BARBIE1 for a full description, but briefly, we set a modern Earth twin as our fiducial data spectrum, free parameters, and priors following \citet{feng18}, with isotropic volume mixing ratios (VMRs) \ce{H2O}$=3\times10^{-3}$, \ce{CH4}$=1.65\times10^{-6}$, \ce{CO2}$=3.84\times10^{-4}$, constant temperature profile at 250 K, $\mathrm{A_{s}}$ of 0.3, $\mathrm{P_{0}}$ of 1 bar, and a planetary radius fixed at $\mathrm{R_p}$ = 1 $\mathrm{R_\Earth}$. The model also approximates a partially cloudy scenario, with the final spectrum comprised of 50\% clear and 50\% cloudy spectra, i.e. $\mathrm{C_{f}}$ = 50\%. $\mathrm{C_{f}}$ is also retrieved on in our simulations. We then run a set of Bayesian spectral retrievals using PSGnest. PSGnest\footnote{https://psg.gsfc.nasa.gov/apps/psgnest.php} is a Bayesian spectral retrieval methodology adapted from the original Fortran version of the Multinest retrieval algorithm \citep{multinest} but adapted for, and housed in, PSG. There were 400 live points used, and an evidence tolerance of 0.1. We then vary the \ce{H2O} abundance following the same values in BARBIE1, while maintaining all of the other parameters to their modern Earth values to isolate the effect of \ce{H2O}, and repeat the retrievals process. Although photometric observations are taken serially, by retrieving on all points simultaneously, the retrieval can perform more accurately with context and break the inherent degeneracies in single-filter observations \citep{ducrot24}.

For this study, we still use the pre-computed grid methodology by S23. We utilize the KEN grid set of terrestrial planet reflected spectra developed and validated in BARBIE3 and stored in PSG. The KEN consists of four geometric albedo spectral grids - Merman, Beach, Lifeguard, and Allan - that contain 7.4 million pre-computed spectra over a variety of molecules, each grid's minimums, maximums, and parameter points described in Figure 2 of BARBIE3. The grids have a native resolving power of R = 500. In this study we utilize the Beach (B-KEN) grid, which contains \ce{H2O}, \ce{CH4}, and \ce{CO2}.

The spectral retrievals process produces several results. We receive the highest-likelihood values, average posterior distribution values, uncertainties, and the log evidence (logZ) \citep{PSGbook} as outputs. We use the logZ to calculate the log-Bayes factor \citep[$\mathrm{lnB}$;][]{benneke13}. $\mathrm{lnB}$ is the most critical component to our analysis, as it directly compares retrievals with and without the molecule of interest to estimate the likelihood of the molecule existing in an exoplanet's atmosphere. It adds an additional layer of understanding to the detection of biosignatures, beyond a simple yes or no, quantifying how much better the model fit is with or without the molecule of interest. We vary our metrics slightly from what is described in \citet[Table 2 of][]{benneke13} wherein $\mathrm{lnB}<2.5$ is unconstrained (no detection) rather than a weak detection, $2.5\le\mathrm{lnB}<5.0$ is a weak detection rather than a moderate detection, and $\mathrm{lnB}{\ge}5.0$ is a strong detection. We also calculate the lower and upper limit values of the 68\% credible region, as well as the median \citep{harrington22}. While this aids in estimating how well-constrained the retrieved value is, $\mathrm{lnB}$ provides a better estimate of detection as it directly investigates the presence of a molecule. Taken together, the 68\% credible region and $\mathrm{lnB}$ can provide the strength and constraint of a detection. 

For the purposes of this work, an unconstrained detection comes with the caveat that it is unconstrained with a normalized exposure time 4x the baseline - the object could be observed longer if a detection is critical.

\subsection{Yield Calculations}

After determining the SNR required for \ce{H2O} detection per R using our grid-based retrieval methodology, we feed those SNR values to a yield calculation code to determine the relative merits of our photometric observing scenarios. 

We use the Altruistic Yield Optimizer (AYO) for yield calculations. AYO optimizes many aspects of the observing strategy, including the selected targets and exposure times, to maximize the yield of the survey. To do so, the AYO algorithm distributes $\simeq10^{5}$ synthetic planets around each star according to the planet type defined by the user (e.g., ``exoEarth candidates"), then calculates their exposure times given an instrument model and background noise sources. The planets are sorted by exposure time, after which the completeness \citep{brown2004} as a function of time is calculated (as well as the derivative dC/dt). The optimal value of dC/dt is then determined for all observations, thus optimizing the selected targets, number of visits per star, and the exposure time for each visit to maximize the resultant yield. AYO also has the capability to optimize the best wavelength for observation on a star-by-star basis, thus allowing for optimizing the bandpass for observation. The HWO Preliminary Input Catalog (HPIC) is used as the input target list \citep{tuchow24}, which contains $\sim$13,000 stars within 50 pc complete to 12$^{\rm th}$ TESS magnitude. For further information on AYO and HPIC, please see \citet{stark14, stark24b}. We will specifically be using the version 17 of AYO implemented into the Framework for Remote Implementation of Demand-based Altruistic Yields (FRIDAY). In this latest version, there is a new degree of observation optimization, wherein the shape and size of the photometric aperture is optimized on a planet-by-planet basis. The shape and size of the PSF is determined via a PSF ``truncation ratio," which sets the portion of the planet's peak-normalized PSF that contributes to the exposure time calculation. The scale of the PSF is treated self-consistently when determining background noise sources and detector noise pixels. 

The high-level astrophysical assumptions are mirrored to \citet[Table 1][]{stark24b}. For our baseline mission, we begin with parameters inspired by Exploratory Analytic Case 1 (EAC1) for the Habitable Worlds Observatory, which assumed an off-axis segmented primary mirror with a 6m inner diameter with two coronagraph channels operating in the VIS and NIR. The VIS channel is assumed to operate from 0.5 - 1 {\microns}, and as we are only interested in the 0.9 {\microns} \ce{H2O} feature, we are therefore interested in utilizing only the visible channel. Starting with the detection criteria, for spectroscopic observations, we assume two parallel coronagraph channels, mimicking two visible coronagraph channels operating in parallel for exoplanet detection. For the photometric observations we conservatively assume one channel, with the observations taken serially, and multiply the resultant exposure time for one spectral point by N, where N is the number of spectral points to mimic this observation structure. The end-to-end throughput of the system is split into multiple terms, one containing all reflectivities and transmissivities, one that budgets for contamination, one that budgets the detector QE, and the coronagraph's core throughout - all vary over wavelength, and are thus set to the value corresponding to $\sim$1 {\microns} for the characterization observations. All throughput values are described in Table~\ref{tab:missionparams}.

Although AYO has the ability to optimize bandpass selection on a star-by-star basis for detection and characterization, we only permit bandpass optimization for detection. The characterization wavelength is set to the long wavelength edge of the 0.94 {\microns} \ce{H2O} feature, as that is the only feature of interest in this study. The characterization wavelength is not allowed to be optimized, as in prior works \cite{stark24a}, and is set to the long wavelength point at continuum as that is a more conservative estimation. For all broadband detections, an SNR of $>$7 is required, following \citet{stark24b}. For spectral characterization, the resolving power varies based on assumed photometric observations (10\%, 5\%, and 3\% bandpasses) or on assumed spectroscopic observations (R = 50, 70, and 140). We also utilize SNRs that result in a strong detection of modern \ce{H2O}, found in the retrievals portion of this study.


We adopt the Amplitude-Apodized Vortex Coronagraph (AAVC, charge 6) designed for EAC1. We place several limits on its assumed performance: we adopt a raw contrast floor of $1\times10^{-10}$ and a minimum working angle of 1$\lambda/D$. Our assumed detector parameters are generic in nature, but are most similar to a Skipper CCD \citep{bebek}, which varies from the intial LUVOIR report. A Skipper has higher QE near 1 {\microns} than, for example, the Roman EMCCD ($\sim$90\% vs $\sim$11.6\%) \citep{teledyne, romanemccd}. We additionally investigate multiple detector noise scenarios, ranging from no noise to a total effective dark current of $1 pix^{-1} s^{-1}$. For most scenarios, we adopt the noise floor (i.e. $\Delta mag_{floor}$) used by \citet{stark25arxiv}; however, we additionally adopt no noise floor for one scenario to investigate the impact of the noise floor on yields. 

Certain parameters must be adapted to account for spectroscopic or photometric observation techniques. For spectroscopic observations, we assume the dispersing instrument will be an integral field spectrograph (IFS), which carries with it additional optics resulting in a 30\% reduction in throughput. For IFS observations, we adopt a conservative 6 pixels per spectral bin per lenslet at the characterization wavelength, since we are not assuming a split for polarization, and the IFS requires an additional multiplicative factor to account for the lenslet array. For a photometric observation simply using an imager, we ignore the throughput penalty of the IFS and adopt one detector pixel for the imaging mode, purely set by the spatial scale of the core of the PSF. These assumptions impact the the spectroscopic yields, making them less ideal, however we are interested in the most conservative approach to yield calculation. Additionally, we assume that the photometric observations are taken serially - we approximate this by calculating the exposure time at one spectral point and multiply that exposure time by N, where N is the number of spectral points rather than individually calculate the exposure time at each spectral point. Additionally, following the LUVOIR study \citep{luvoir}, we mandate six initial visits to every target, under the assumption that this will result in approximately 3 detections of a planet, necessary for confirmation and orbit determination. Table~\ref{tab:missionparams} summarizes the mission parameter assumptions given to AYO for this study, along with details on the parameters differed for spectroscopic and photometric observations.

\begin{table*}[h!]
\hspace{-2cm}
\begin{threeparttable}
\caption{Coronagraph Mission Parameters}
\begin{tabular}{ccc}
            \hline
            \hline
            \textbf{Parameter} & \textbf{Value} & \textbf{Description}\\
            \hline
             & & \textbf{General Parameters}\\
            $\sum_\tau$ & 2 years & Total exoplanet science time out of an assumed 5 year mission\\
            $\tau_{dynamic}$ & 1.1 & Multiplier overhead to touch up dark hole\\
            $\tau_{static}$ & 2.32 hrs & Overhead for slew, settling, and digging dark hole\\
            $X$ & optimized per planet & Inscribed photometric aperture radius to extract planet signal ($\lambda/D_{LS}$)\\ 
            $\varsigma_{floor}$ & $1\times10^{-10}$ & Raw contrast floor \\
            Post-Processing Factor (PPF) & [30, 10,000] & Setting the 1$\sigma$ noise floor contrast (noise floor = $\varsigma_{floor}$/PPF) \\
            $T_{contam}$ & 0.95 & Effective throughput as driven by contamination \\
            \hline
             & & \textbf{Detection Parameters}\\
            $\lambda_{d}$ & 650 $nm^{a}$ & Central wavelength for detection \\
            $SNR_{d}$ & 7 & Required SNR for detection \\
            $T_{optical}$ & 0.338$^{a}$ & End-to-end reflectivity at $\lambda_{d}$ \\
            $\tau_{d,limit}$ & 2 months & Detection time limit, overheads included \\
            \hline
             & & \textbf{Characterization Parameters}\\
            $\lambda_{c}$ & $\sim$1000 nm & Long wavelength edge for characterization \\
            $SNR_{c}$ & [10.02, 8.46, 5]$^{b}$ & SNR required for characterization with IFS \\
            & [15.59, 14.7, 14]$^{c}$ & SNR required for characterization with photometry\\
            $R$ & 50, 70, 140 & Spectral resolving power for IFS \\
             & 10\%, 5\%, 3\% & bandwidth for photometry\\
            $T_{optical,IFS}$ & 0.233 & End-to-end reflectivity at $\lambda_{c}$ for IFS\\
             & 0.332 & End-to-end reflectivity at $\lambda_{c}$ for photometry\\
            $\tau_{c,limit}$ & 2 months & Characterization time limit, overheads included \\
            \hline
             & & \textbf{Detector Parameters}\\
            $n_{pix,d}$ & 1 & Number of pixels in photometric aperture per imager at $\lambda_{d}$\\
            $n_{pix,c}$ & 6 & Number of pixels per spectral bin in coronagraph IFS at $\lambda_{c}$ \\ 
             & 1 & Number of pixels in photometric aperture per imager \\ 
            Total Effective Noise Rate & [0, $1\times10^{-6, 4, 3, 2, 1, 0}$]$^{d} pix^{-1} s^{-1}$ & $\xi+RN+\tau_{read}+CIC^{e}$ \\ 
            $T_{QE}$ & 0.9 & Raw detector QE at all wavelengths \\
            $T_{dQE}$ & 1 & Effective throughput due to bad pixel/cosmic ray mitigation \\
            $T_{core}$ & 0.366179 & Core Coronagraph Throughput \\ 
            \hline
            \hline
\end{tabular}
\begin{tablenotes}
\item $^{a}$Values for the most likely bandpass, however AYO optimizes bandpass and will adjust.
\hspace{0.35cm} \item $^{b}$SNRs for strong detections of \ce{H2O} at R = 50, 70, 140, respectively.
\item  $^{c}$SNRs for strong detections of \ce{H2O} at 10\%, 5\%, 3\% bandwidth , respectively.
\item  $^{d}$No noise, and multiple iterations of noise cases.
\item  $^{e}$Encompassing dark current ($\xi$), read noise (RN), time between reads ($\tau_{read}$), and clock induced charge (CIC).
\end{tablenotes}
\label{tab:missionparams}
\end{threeparttable}
\end{table*}

\section{Results}
\label{sec:results}

\begin{figure*}
\centering
\gridline{\fig{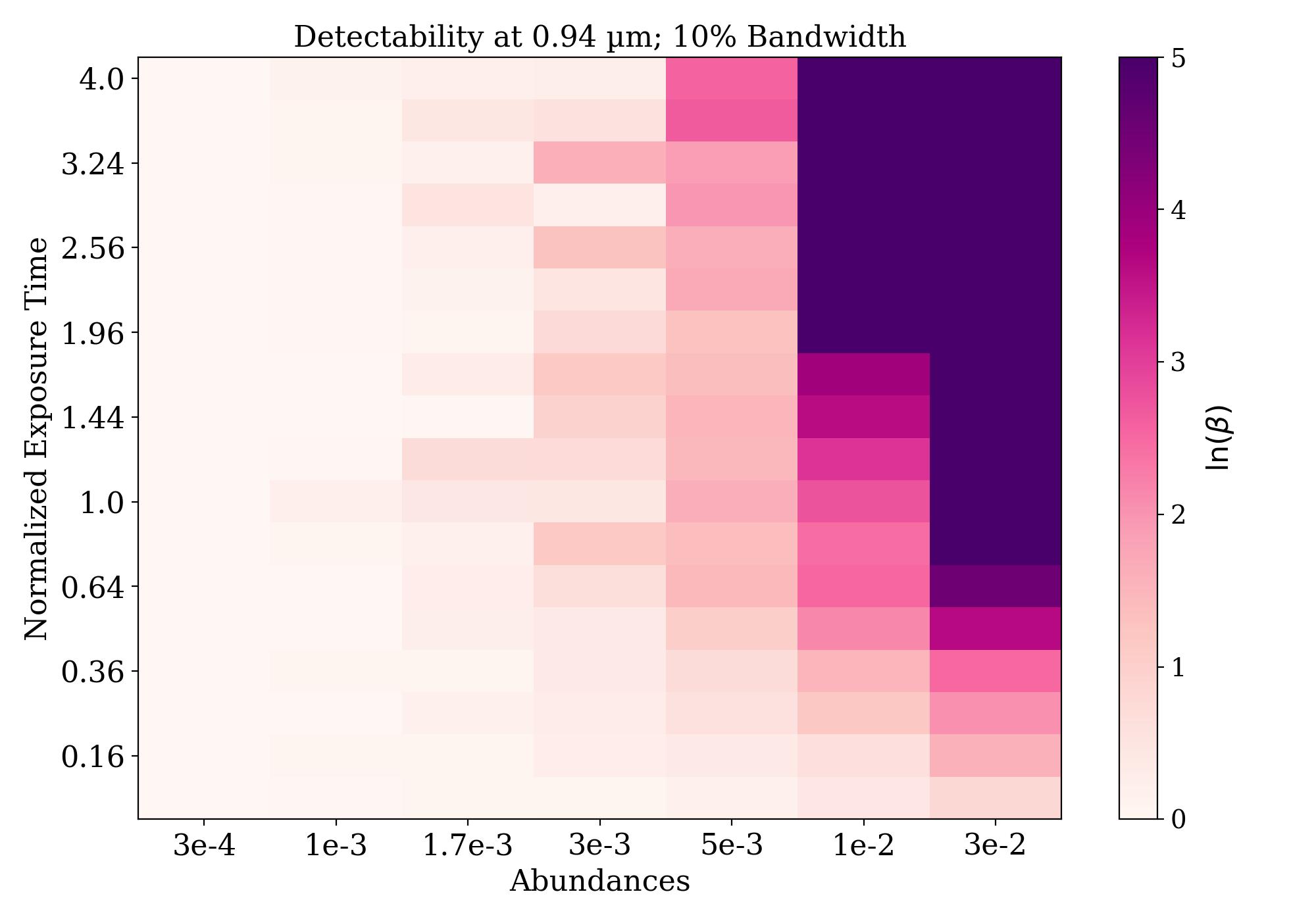}{0.334\textwidth}{(a)}
          \fig{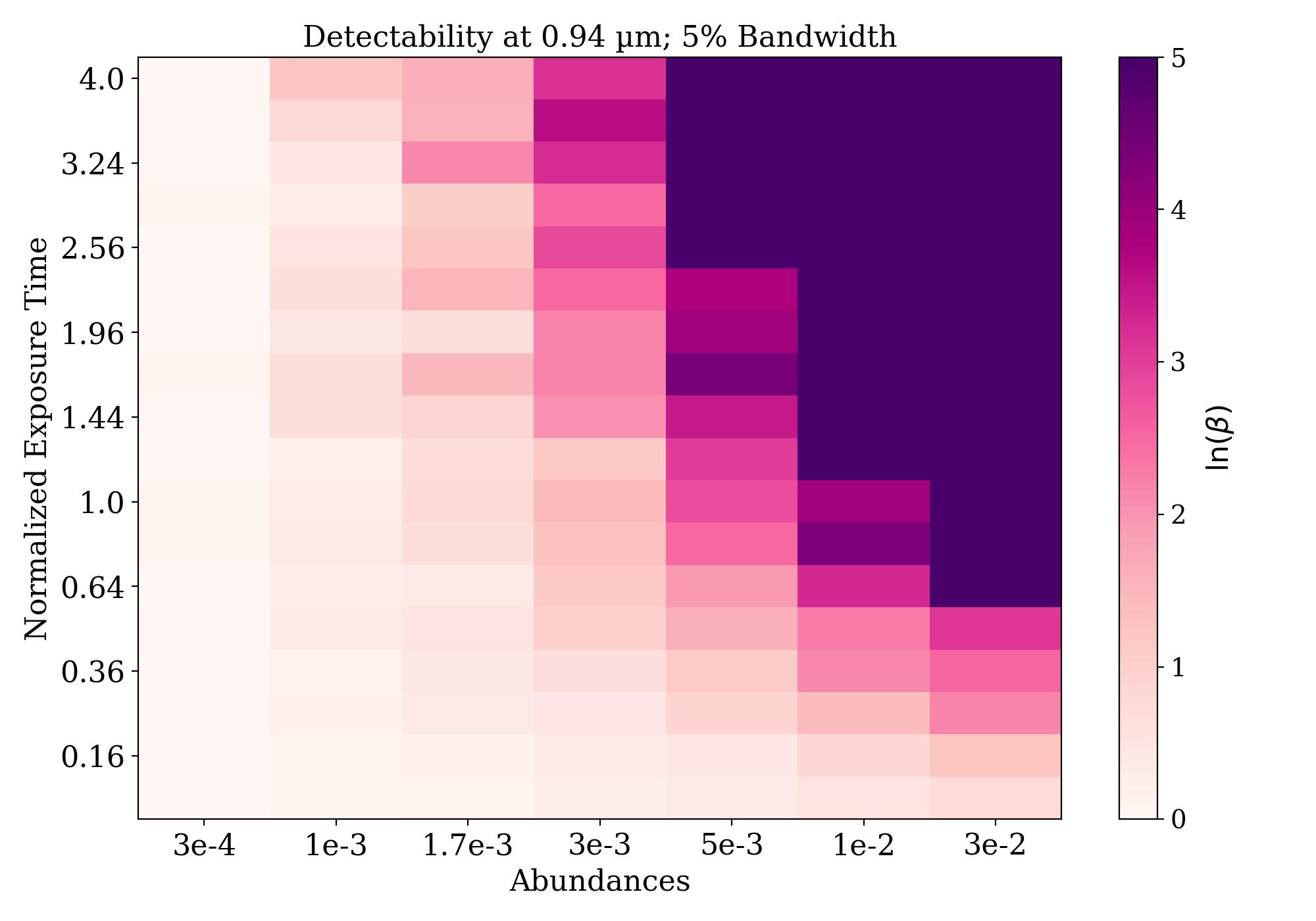}{0.334\textwidth}{(b)}
          \fig{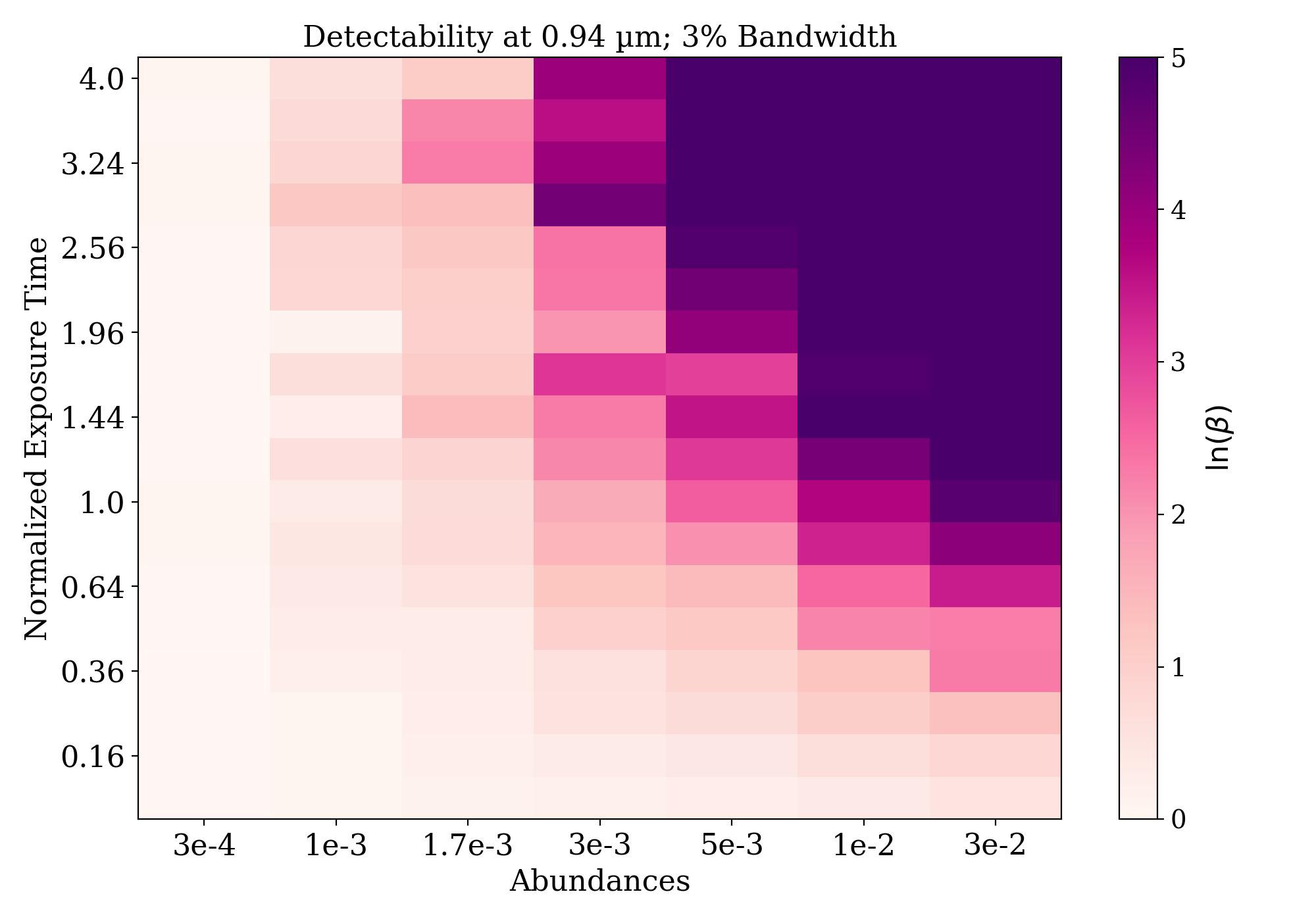}{0.334\textwidth}{(c)}}
\caption{Heat map plots illustrating detection strength as a function of normalized exposure time and varying resolving abundance at varying Earth \ce{H2O} abundances with 2 spectral points permitted at each bandwidth. $\mathrm{lnB}$$\ge$5 is a strong detection, 2.5$\le$$\mathrm{lnB}$$\le$5 is a weak detection, and $\mathrm{lnB}$$\le$2.5 is unconstrained. A modern abundance of \ce{H2O} is considered to be $3\times10^{-3}$. \textbf{With two spectral points regardless of bandwidth, \ce{H2O} detectability is low, requiring higher abundances.}}
\label{fig:heatmap2pt}
\end{figure*}

We begin with a test of 2 spectral points, placed as seen in Figure~\ref{fig:fullspec_rcomp} upper panel. In Figure~\ref{fig:heatmap2pt}, we present a series of heat maps showing the detection strength as a function of \ce{H2O} abundance (x-axis) and normalized exposure time (y-axis) for the two-point analysis. From left to right, we show bandpass widths of 10\%, 5\%, and 3\%, respectively. A 20\% bandwidth is not explored, as the photometric bandwidths would overlap with each other and this is not desired for this work. The color bar represent a range of $\mathrm{lnB}$ between unconstrained to strong detection (0-5). We see that at a bandwidth of 10\% with 2 photometric points (Figure~\ref{fig:heatmap2pt}a), strongly detecting \ce{H2O} would require abundances an order of magnitude higher than modern Earth values, and otherwise remain unconstrained with an exposure-time limit of 4x the baseline. In Figure~\ref{fig:heatmap2pt}b, at 5\% bandwidth with two spectral points, abundances less than $5\times10^{-3}$ VMR are not strongly detectable (including a modern Earth abundance), and indeed even at $5\times10^{-3}$ VMR a normalized exposure time of 2.56 is required for strong detection. The normalized exposure time requirements for higher abundances drops, with the highest abundance tested ($1\times10^{-2}$ VMR) requiring an normalized exposure time of 0.81 for strong detection. A modern Earth abundance is able to be weakly detected, but only at normalized exposure time$\ge$2.89. Figure~\ref{fig:heatmap2pt}c, at 3\% bandwidth with two spectral points, the results are nearly identical to Figure~\ref{fig:heatmap2pt}b. Indeed, the only notable differences in the results are that, although a modern Earth abundance is still only weakly detectable, the weakness of the detection is lesser than that at 5\% at the same normalized exposure time, and the highest tested abundance of $1\times10^{-2}$ VMR also fares worse, requiring an normalized exposure time of 1.21 for strong detection rather than an normalized exposure time of 0.81 as for 5\% bandwidth. This is likely due to the narrowness of the bin at 3\% compared to 5\%.

\begin{figure*}
\centering
\gridline{\fig{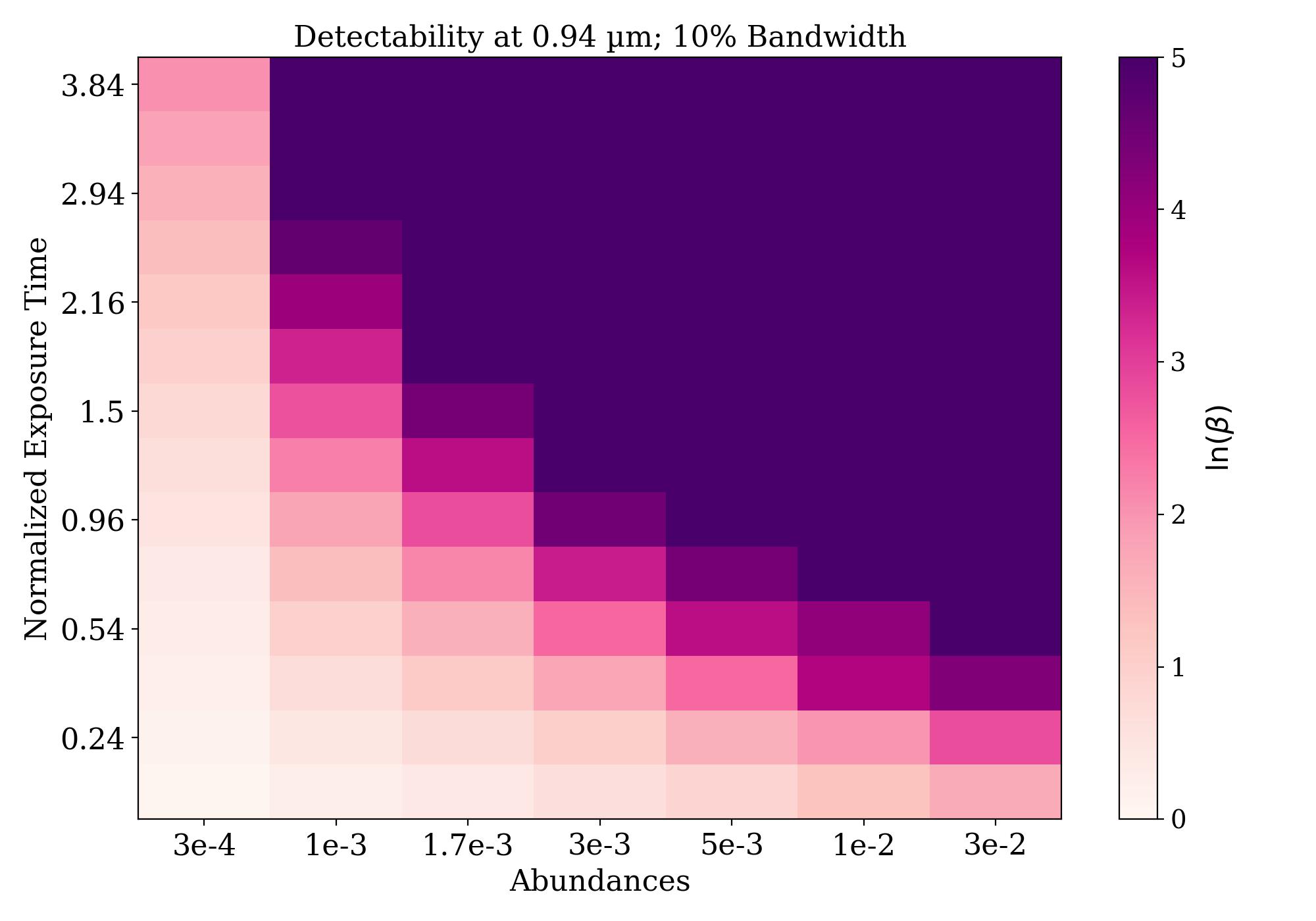}{0.334\textwidth}{(a)}
          \fig{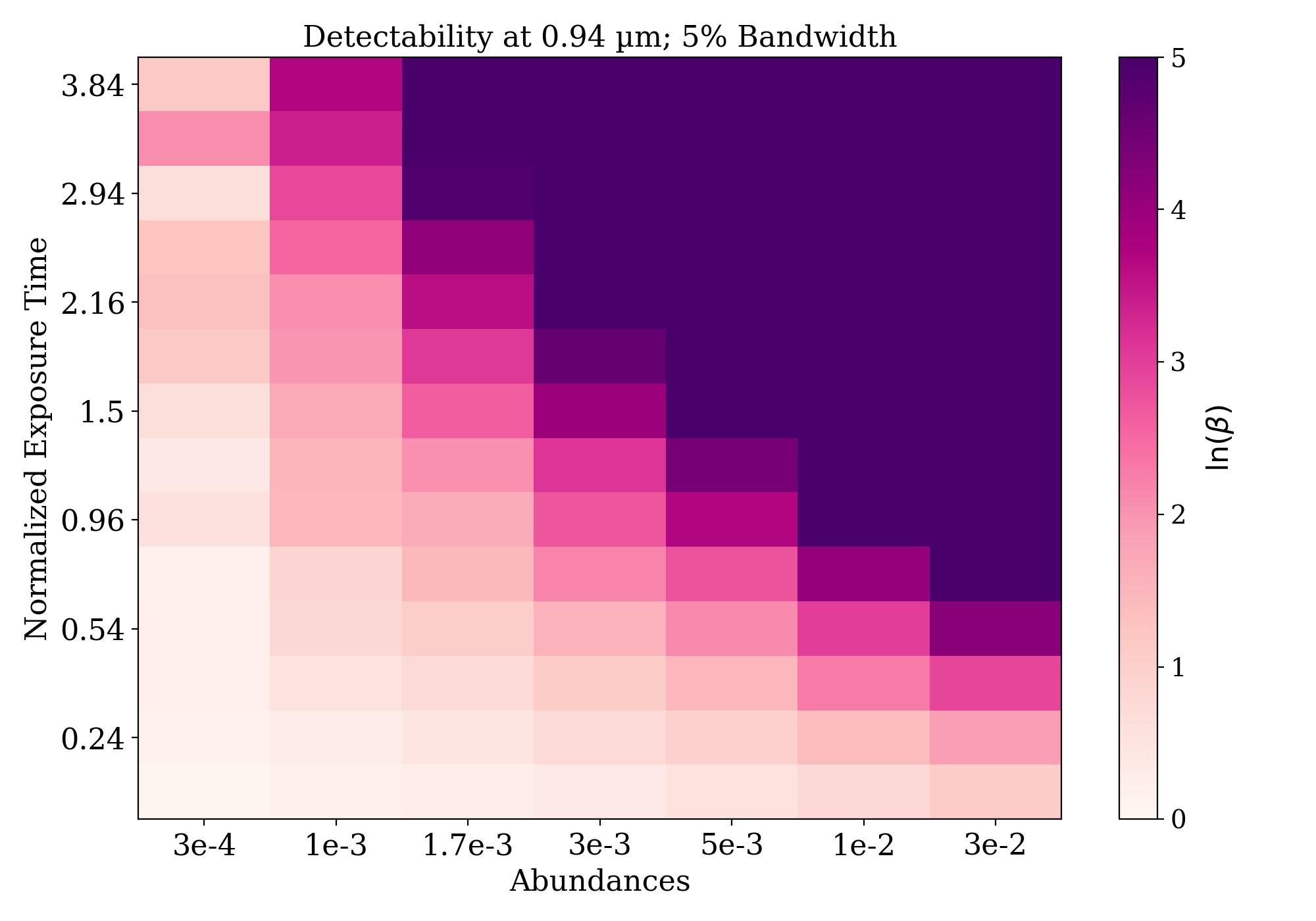}{0.334\textwidth}{(b)}
          \fig{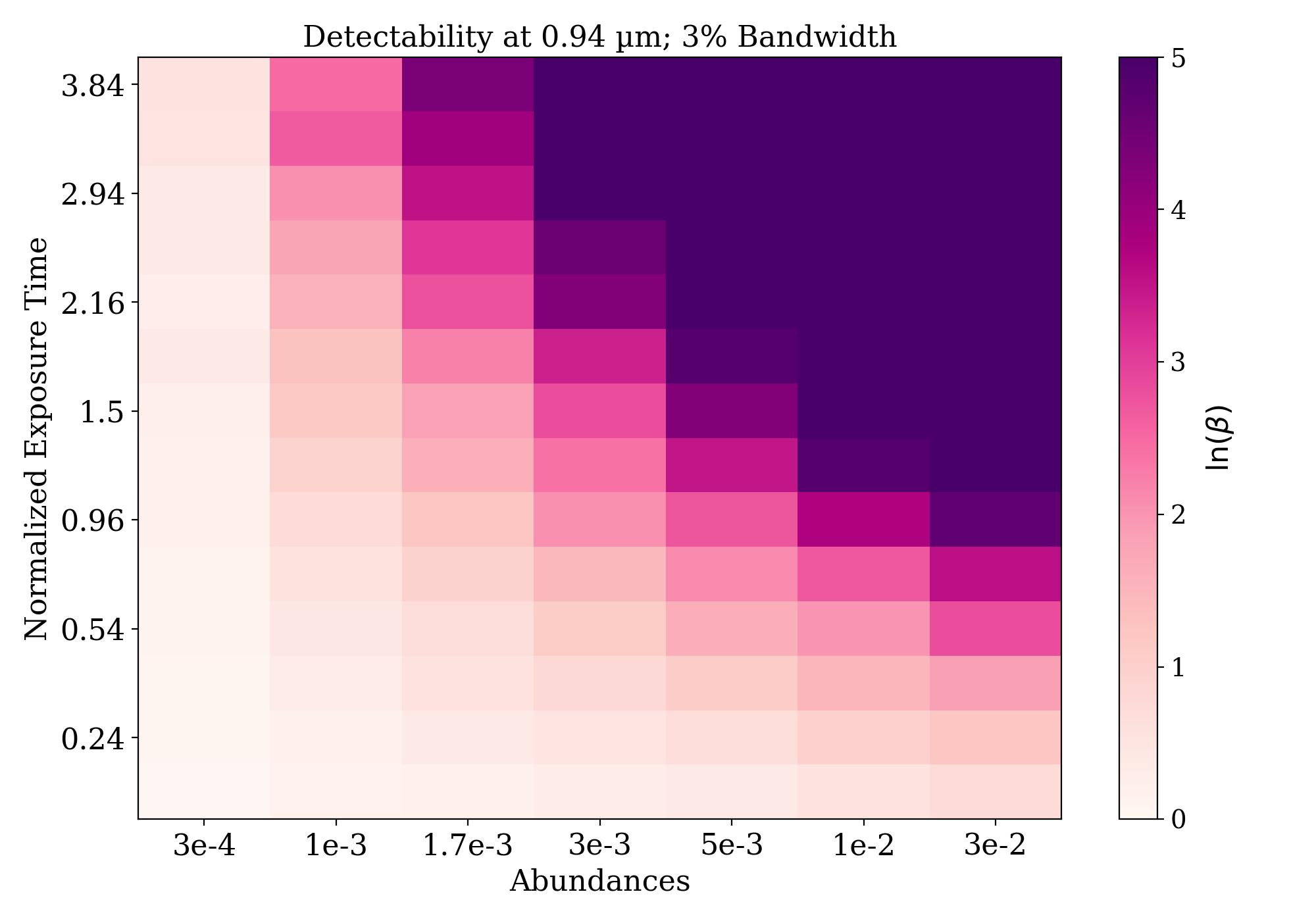}{0.334\textwidth}{(c)}}
\caption{Heat map plots illustrating detection strength as a function of normalized exposure time and varying resolving abundance at varying Earth \ce{H2O} abundances with 3 spectral points permitted at each bandwidth. $\mathrm{lnB}$$\ge$5 is a strong detection, 2.5$\le$$\mathrm{lnB}$$\le$5 is a weak detection, and $\mathrm{lnB}$$\le$2.5 is unconstrained. Note that the y-axis is cropped to 3.84, to more directly compare to Figure~\ref{fig:heatmap2pt} which caps at 4.0. \textbf{The key factor to photometric \ce{H2O} detection is spectral point quantity.}}
\label{fig:heatmap3pt}
\end{figure*}

We next move to investigating 3 spectral points permitted per R, the placement of which is shown in Figure~\ref{fig:fullspec_rcomp} bottom panel. In Figure~\ref{fig:heatmap3pt}, we present another series of heat maps showing the detection strength as a function of \ce{H2O} abundance (x-axis) and normalized exposure time (y-axis) for the three point analysis. From left to right, we show bandwidths of 10\%, 5\%, and 3\%, respectively. We see an incredibly stark difference in results - looking at all of the panels of Figure~\ref{fig:heatmap3pt} simultaneously, we see that a 10\% bandwidth yields the most promising results, followed by 5\%, and lastly, 3\%. In Figure~\ref{fig:heatmap3pt}a, at 10\% bandwidth, all abundances except for the lowest one are strongly detectable, at moderate to low normalized exposure times. At the lowest detectable abundance ($1\times10^{-3}$ VMR), an normalized exposure time of 2.94 is required for a strong detection. At a modern Earth abundance, an normalized exposure time of 1.22 is required for strong detection. In Figure~\ref{fig:heatmap3pt}b, at 5\% bandwidth, we see worsening results; $1\times10^{-3}$ VMR can now only achieve a weak detection at best, with a normalized exposure time of 3.84, and a modern Earth requires an normalized exposure time of 2.16 for strong detection. Moving to Figure~\ref{fig:heatmap3pt}c, 3\% bandwidth, we find that $1\times10^{-3}$ VMR is now no longer strongly detectable at any normalized exposure time$\le$3.84, and a modern Earth requires an normalized exposure time of 2.94 for strong detection. Again, this is likely due to the narrowing bin as the resolving power is increased. 

\subsection{AYO Exoplanet Yield Calculations}

To understand the impact of these results on scientific productivity, we perform exoplanet yield calculations using AYO. Only the three point photometric results were investigated, due to the significantly lower detectability of the two point photometric scenario. We will assume that a minimum of three points is required for strong detection of \ce{H2O}. We investigate multiple noise cases, as we expect the noise to be the strongest driver in the benefit of photometric observations. We also constrain the characterization wavelength to the longest point in the studied bandpasses, the approximation of which is shown in Table~\ref{tab:missionparams}. Additionally, we investigate the three bandwidths that correspond to photometric observations (3\%, 5\%, 10\%) and three resolving powers that correspond to spectroscopic observations (R = 50, 70, 140). All spectra were developed using PSG, in the same fashion described in prior BARBIE works \citep{latouf23, barbie3}. 

\begin{figure*}[ht!]
\centering
\includegraphics[scale=0.5]{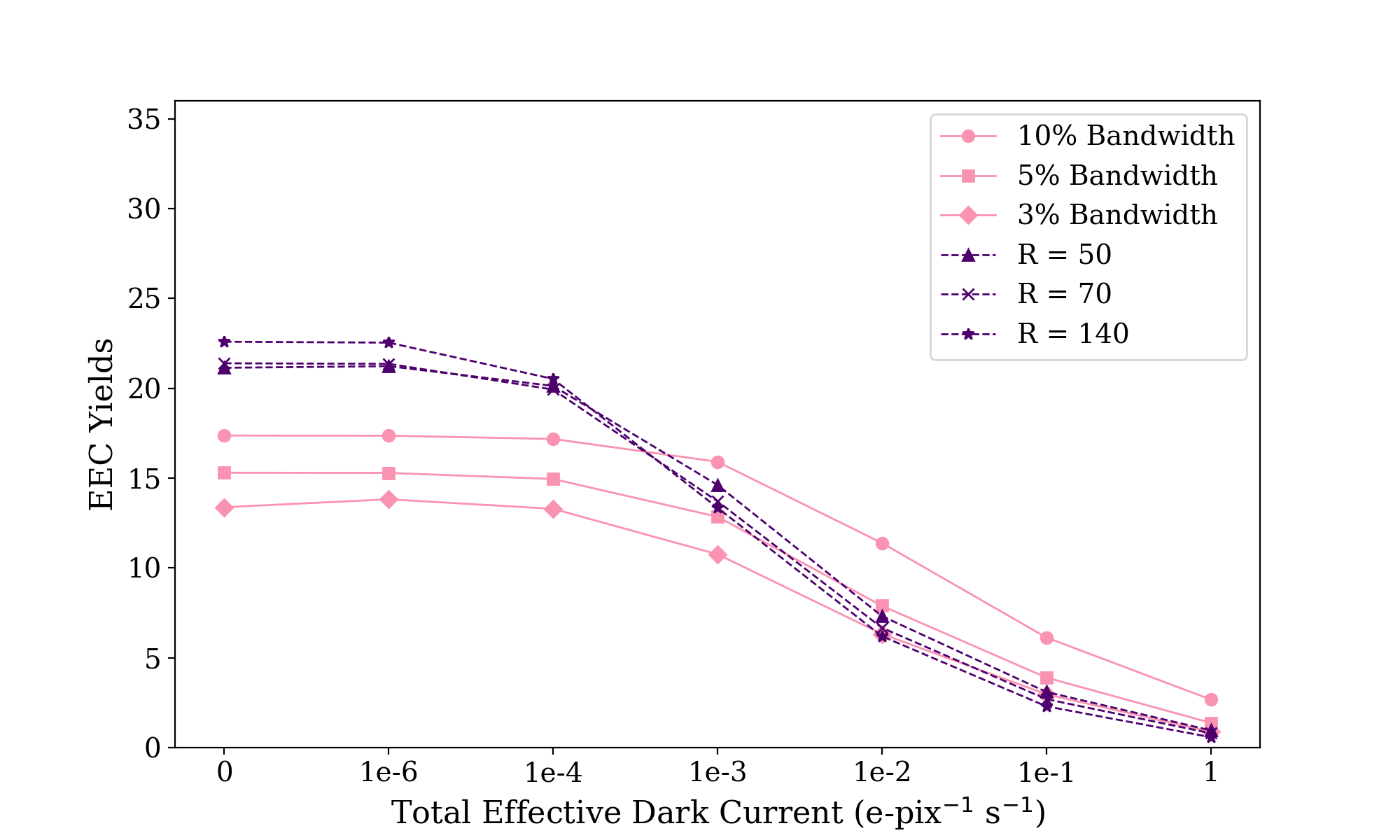}
\caption{ExoEarth candidate (EEC) yields for three photometric scenarios (solid pink lines) and three spectroscopic scenarios (dashed purple lines) as a function of detector noise. \textbf{Spectral observations result in higher yields at low noise cases, but photometric observations result in higher yields at higher noise cases.}}
\label{fig:yields_long}
\end{figure*}

Figure~\ref{fig:yields_long} shows the resultant exoEarth candidate (EEC) yields for every scenario as a function of detector noise, assuming a three-point photometric scenario, as those yielded the most promising detectability results (i.e. more detections at lower abundances). We find that at very low values of detector noise (i.e. $1\times10^{-6} pix^{-1} s^{-1}$) spectroscopic observations result in greater EEC yields, with R = 140 achieving the highest number, with R = 50 and 70 nearly indistinguishable. This is in spite of the assumption that the IFS has 30\% lower throughput and is likely a result of assuming that photometric observations are taken serially. Although the photometric observations still result in moderate to high yields, with higher yields at broader bandwidths, spectroscopy remains the preferred observational method. As the noise increases, the spectroscopic yields begin to drop significantly, with a broadband photometric observation (10\% bandwidth) resulting in higher yields than any spectroscopic observation at a noise level of $1\times10^{-3} pix^{-1} s^{-1}$. However, spectroscopic observations still have better yields than photometric observations with narrower bands, such as 5\% or 3\%, at this noise instance. As the detector noise continues to increase, the spectroscopic observation yields drop precipitously, with 10\% bandwidth photometry consistently resulting in the highest yields.


\begin{figure*}
\centering
\gridline{\fig{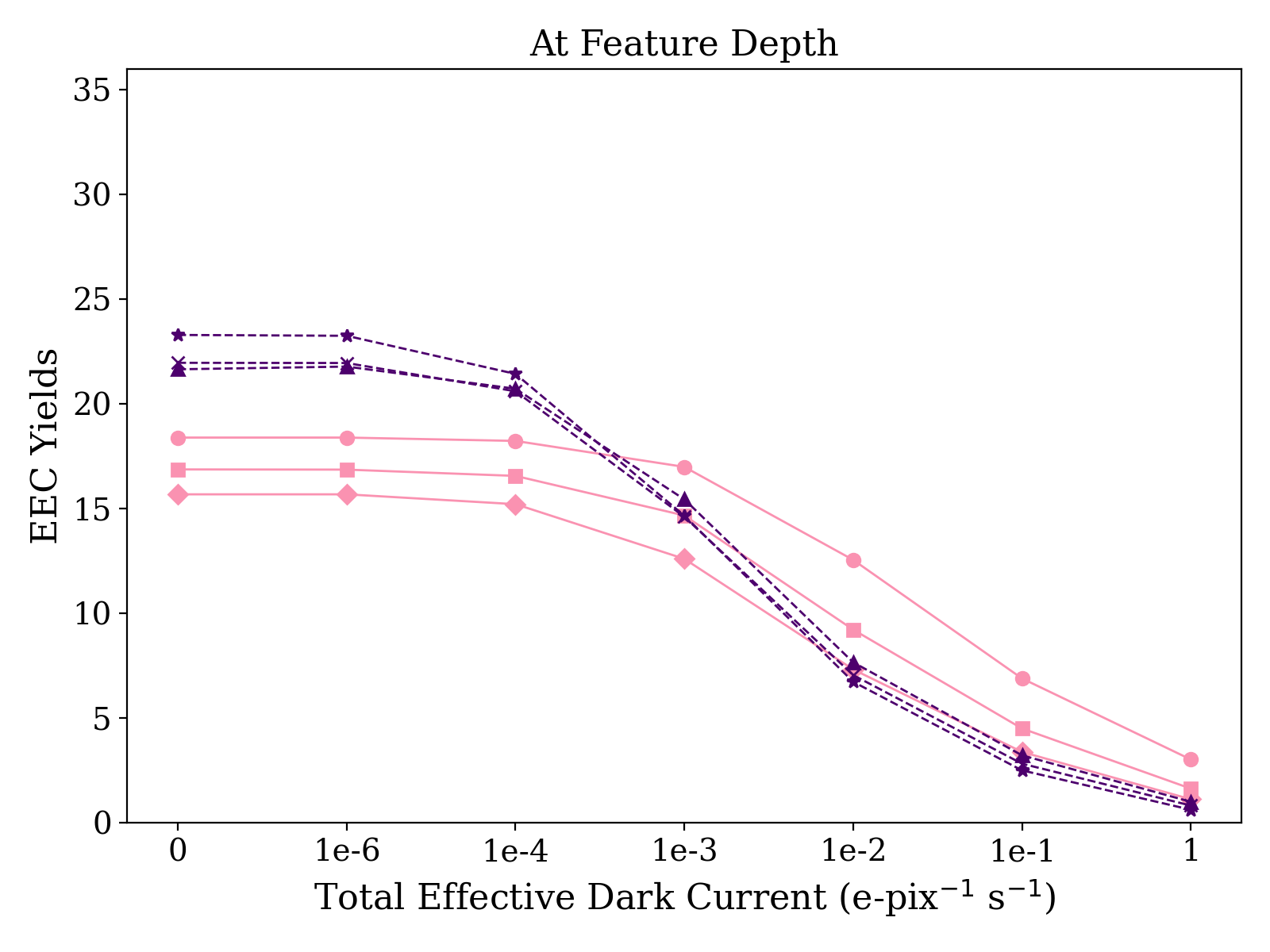}{0.334\textwidth}{(a)}
          \fig{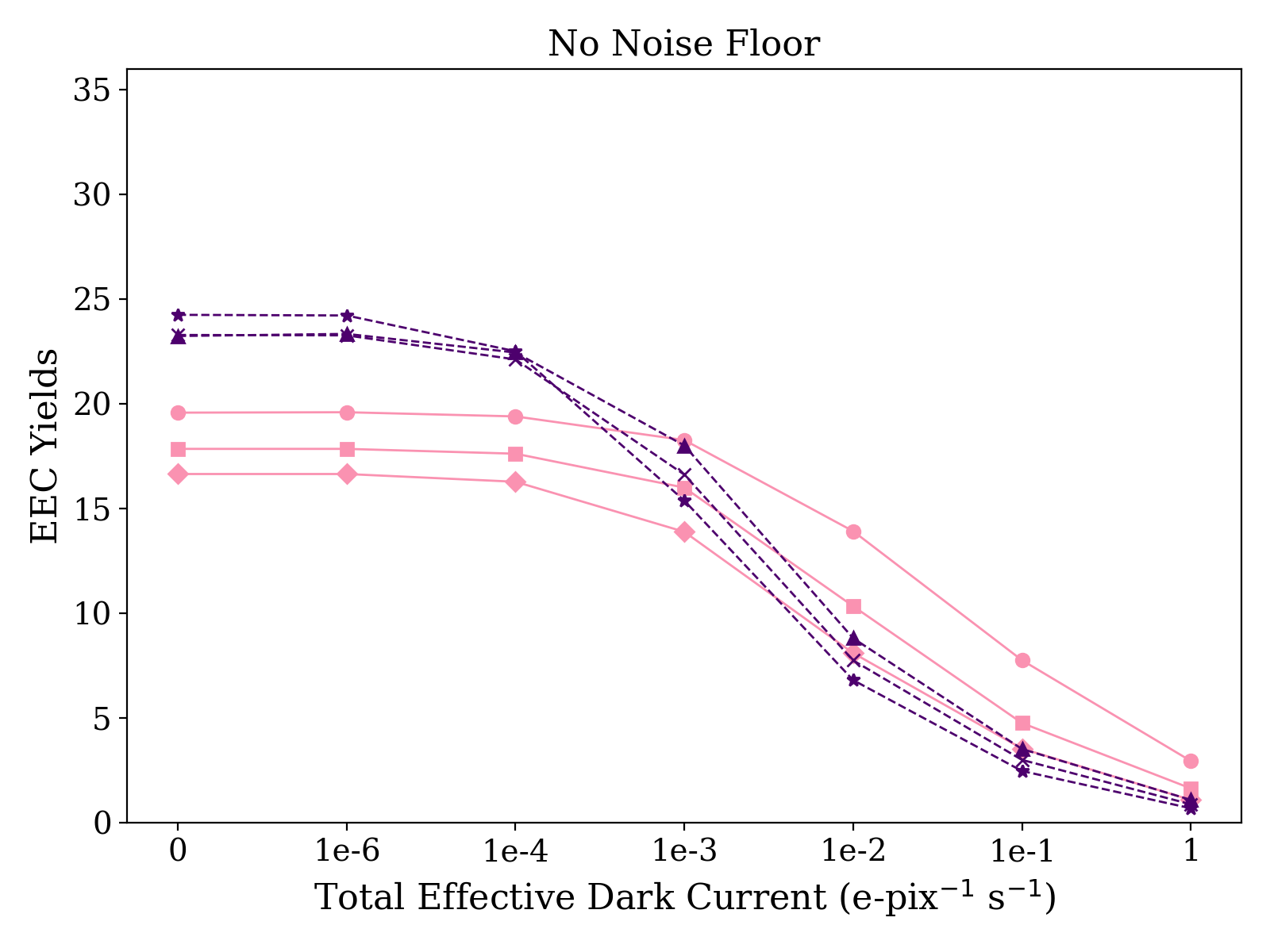}{0.334\textwidth}{(b)}
          \fig{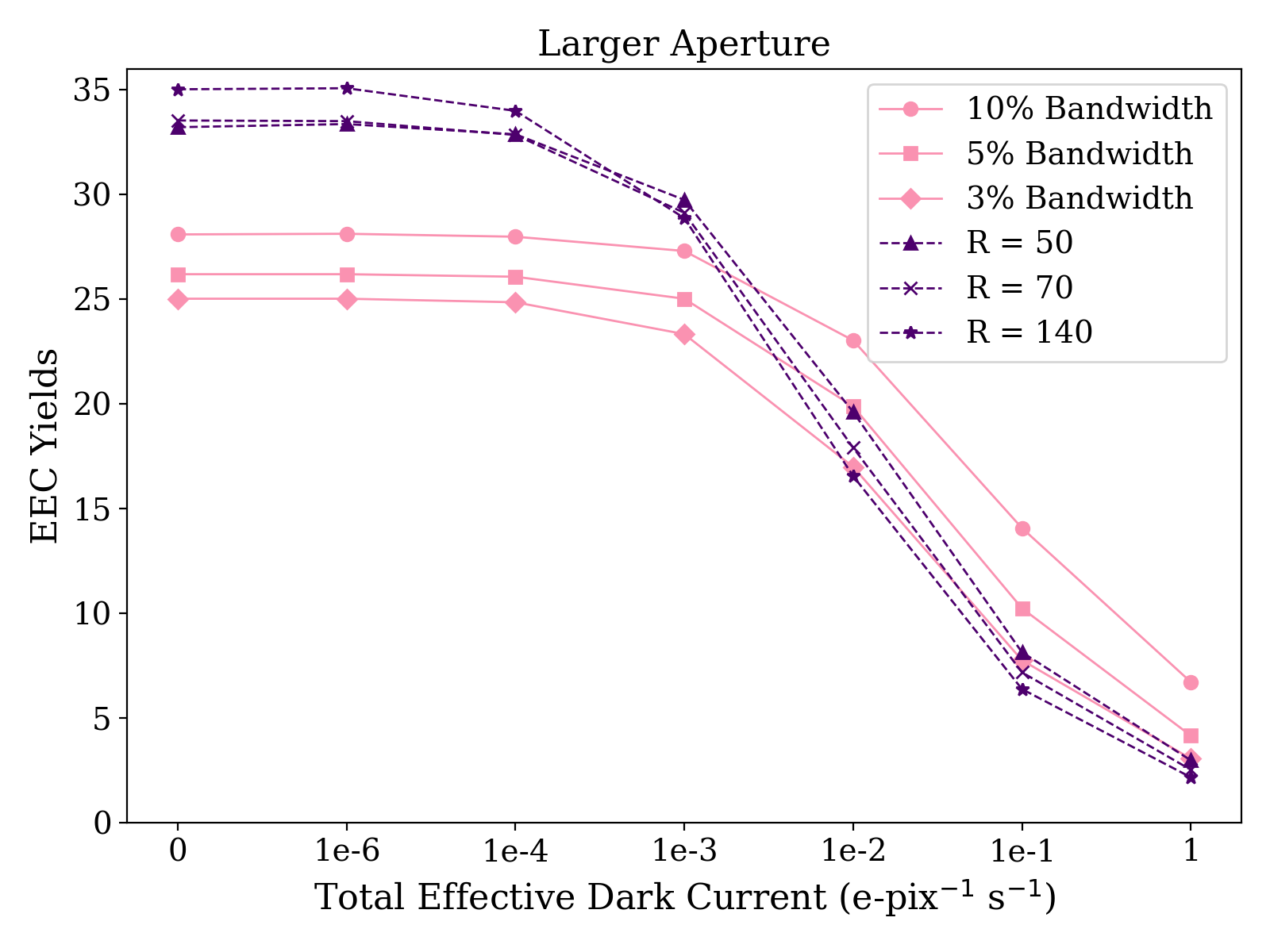}{0.334\textwidth}{(c)}}
\caption{We present the exoEarth candidate (EEC) yields (y-axis) for 3 photometric bandwidths (10\%, 5\%, 3\%) and 3 spectroscopic resolving powers (R = 50, 70, 140) over multiple different noise cases (x-axis), with other simulation variations. The left panel (a) has moved the detection wavelength to the depth of the feature vs the long wavelength point shown previously. The center panel (b) has no noise floor limitation. The right panel (c) has an increased aperture size, from a 6 meters to 8 meters inscribed. All other aspects are identical to Figure~\ref{fig:yields_long}. \textbf{Characterization wavelength placement and noise floor are not the main drivers in photometric vs spectroscopic observation preference - aperture contributes heavily, but the noise is the main decider in preferred observational method.}}
\label{fig:yields_mult}
\end{figure*}

In order to isolate if another factor could be driving these curves in addition to (or even more than) the detector noise, we ran other simulations with variations to investigate yield changes. Figure~\ref{fig:yields_mult} shows the resultant EEC yields for every resolving power as a function of detector noise, but with changes made to the instrument/observation assumptions: Figure~\ref{fig:yields_mult}a moves the characterization wavelength to the spectral point at the depth of the feature which is more optimistic rather than the previously-used, conservative long wavelength point; Figure~\ref{fig:yields_mult}b has no noise floor imposed on the yield calculation, but still selecting the longer wavelength spectral point; and Figure~\ref{fig:yields_mult}c increased the aperture size from a 6-meter inscribed circular aperture size to 8 meters. 

We find in Figure~\ref{fig:yields_mult}a that there is very little difference in resultant yields by selecting the spectral point at the depth of the feature for our exposure-time calculations instead of the longer wavelength spectral point, with only a $\sim$5.5\%, $\sim$6.5\%, and $\sim$10\% increase in yields for R = 50, 70, and 140 respectively. These increases are not significant enough to necessarily justify the more expensive spectroscopic observation. This also indicates that minutia in the wavelength selected for characterization is not a main driver of EEC yields. 

For the results in Figure~\ref{fig:yields_mult}b we increased the noise floor post-processing factor (PPF) to 10,000, which effectively eliminates the noise floor altogether, and we find subtle differences in the resultant yields comparing to Figure~\ref{fig:yields_long}. We find that spectroscopic observations at R = 50 and R = 70 have significantly higher yields at higher noise instances, such as a $\sim$23\% and $\sim$21\% increase respectively at $1\times10^{-3} pix^{-1} s^{-1}$. At lower R, the IFS has lower noise due to collecting more photons per resolution element - however, generally lower Rs require higher SNR for molecular detection, and the noise floor prevents achieving the SNR. By lowering the noise floor, the SNR required for molecular detection is more readily accessible, and thus the lower noise floor improves these results more significantly. However, they still do not perform better than a photometric observation at 10\% width. R = 140 also improves in resultant yields, with an increase of $\sim$15\%, but this still pales in comparison to the increases at lower resolutions and does not compete with the yields expected at photometric observations.

In Figure~\ref{fig:yields_mult}c, we find far more differences than either of the other panels. We find that every observing method results in significantly increased yields when the aperture size is increased, the reasons for which are discussed further below. Spectroscopic observations at R = 140 increase in yields by a whopping $\sim$116\%, with R = 70 and R = 50 increasing by $\sim$112\% and $\sim$103\% respectively. Additionally, we find that photometry no longer results in higher yields than spectroscopy at the $1\times10^{-3} pix^{-1} s^{-1}$ noise case. Spectroscopy at every resolving power results in higher yields, thereby pushing the noise case at which photometry is preferred to higher values, at $1\times10^{-2} pix^{-1} s^{-1}$ and above.


\section{Discussion}
\label{sec:disc}

\begin{figure*}[ht!]
\centering
\includegraphics[scale=0.25]{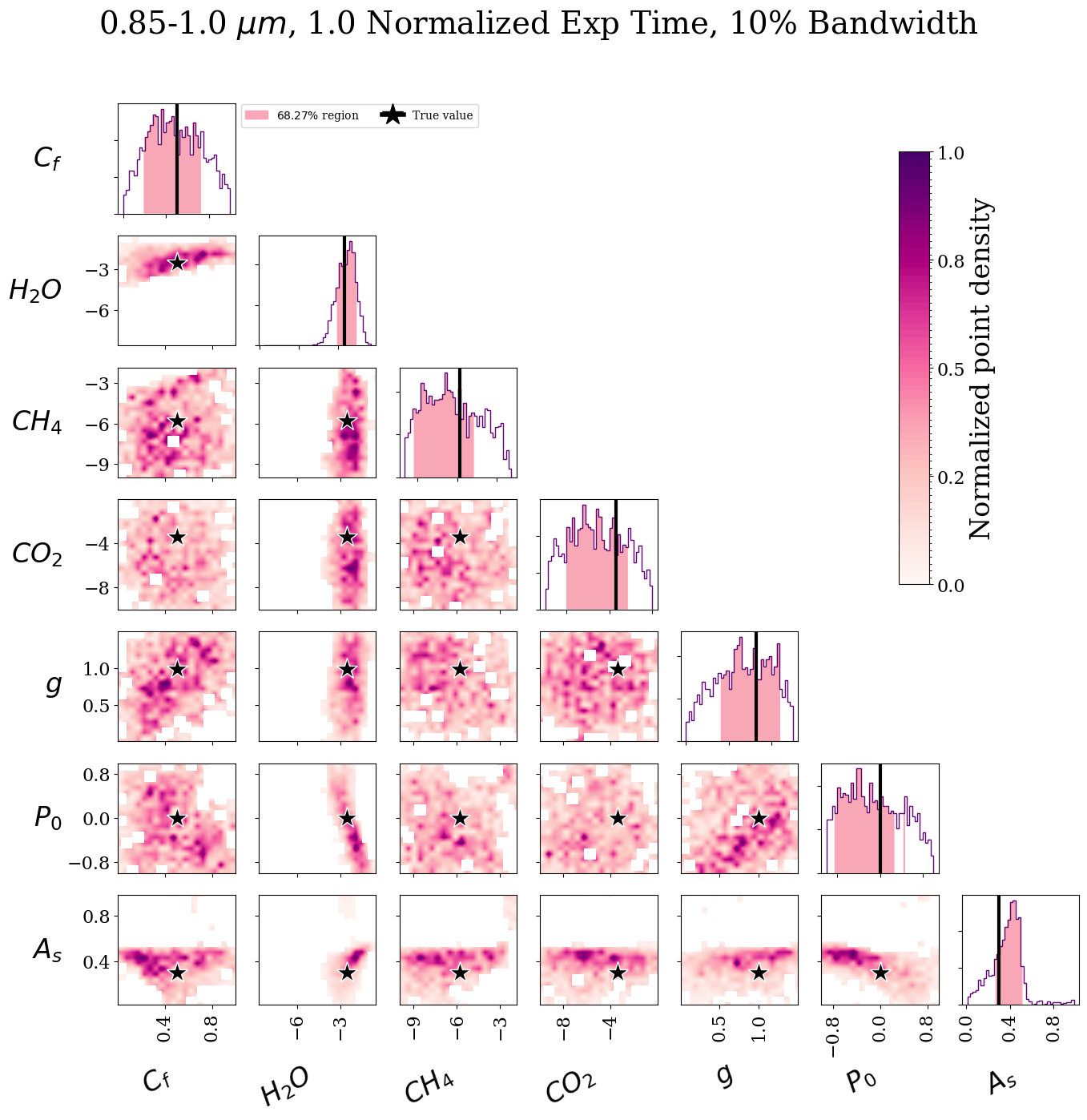}
\caption{Corner plot for the 10\% photometric bandwidth investigated in our study with 3 spectral points, for a modern Earth abundance, $3\times10^{-3}$ VMR. The corner plot contains the posteriors of the entire wavelength region investigated, 0.85 - 1.05 {\microns}. The 68\% credible regions are shown as pink shading in the 1D marginalized posterior distributions along the diagonal of the corner plot, and the true values are represented by black lines in the diagonals of the corner plot, and black stars within the 2D plots. \textbf{Although the posteriors can appear constrained, it does not indicate a strong detection.}}
\label{fig:corner}
\end{figure*}

Understanding the possibility of photometric detection of atmospheric constituents, specifically biosignatures, is an important factor is optimizing the required telescope observing time. However, by reducing the resolving power and thus the amount of spectral points observed, other degeneracies are induced that are not found with IFS observation. For instance, Figure~\ref{fig:fullspec_rcomp} shows a critical difference in geometric albedo from 10\% to 3\% bandwidth. At narrower bandwidths, the albedo is not accurately understood at the deepest absorption point in the feature, leading to higher albedo and thus an artificially more shallow \ce{H2O} feature. The albedo varies as a function of resolving power due to the wavelength binning. As the width of the spectral range covered by each spectral bin increases (i.e. as the resolving power is lowered), greater amounts of the spectral feature edges are included in each bin, which then directly dilutes the spectral depth in each bin. This results in a spectral point with diluted depth, as we see in Figure~\ref{fig:fullspec_rcomp} for 10\% bandwidth. 

Specifically, although the albedo at the depth of the \ce{H2O} feature is approximately 0.12, the feature depth point at 10\% bandwidth is at an albedo of 0.2, which is significantly brighter. At 10\% bandwidth, as shown in Figure~\ref{fig:corner}, this albedo inaccuracy results in an over-retrieved value of $\mathrm{A_{s}}$, a seemingly well-constrained \ce{H2O} abundance, but more critically as shown in Figure~\ref{fig:heatmap2pt}a, in non-detectable modern-Earth \ce{H2O} at this broad bandwidth, requiring higher \ce{H2O} abundances to achieve a strong detection. This is likely due to the fact that as you deepen the feature depth (i.e. increase abundance), although the albedo remains brighter at the feature depth than is the true value, it still decreases, and thus is probing a deeper feature due to the blending from the wings and continuum. As you decrease the bandwidth to 5\% and then 3\%, the albedo gradually dims, approaching the true albedo at the feature depth, as seen in Figure~\ref{fig:fullspec_rcomp}. However, the photometric point placement and number of points are the more powerful drivers for \ce{H2O} detection. 

We find that with two photometric points, only high \ce{H2O} abundances are strongly detectable at any bandwidth, as shown in Figure~\ref{fig:heatmap2pt}. At 5\% (\ref{fig:heatmap2pt}b) and 3\% (\ref{fig:heatmap2pt}c) bandwidths, higher abundances of \ce{H2O} are strongly detectable where that is not possible at a bandwidth of 10\% (\ref{fig:heatmap2pt}a), but at varying normalized exposure time. For 5\% bandwidth, a normalized exposure time of 2.56 is required, where at a 3\% bandwidth, a normalized exposure time of 2.89 is required for strong detection. Detectability at narrower bandwidths is more accessible, due to the position of the feature depth spectral point - at narrower bandwidths, the albedo at the feature depth is dimmer, more accurate to the true value. This leads to a better constraint on the feature, and thus higher detectability at lower \ce{H2O} abundances. However, a modern Earth abundance is not detectable at any normalized exposure time$\le$4.0 for any bandwidth. 

Comparatively, when we investigate having three photometric points, the difference for a 10\% bandwidth is staggering. In Figure~\ref{fig:heatmap3pt}a, all abundances except the lowest tested abundance are detectable, at varying normalized exposure times. For the 5\% bandwidth, although the same abundances are detectable as for 10\% bandwidth, the normalized exposure time required for strong detection increases for every abundance. For 3\% bandwidth, the lowest detectable abundance increases from $1\times10^{-3}$ VMR to $1.7\times10^{-3}$ VMR and again higher normalized exposure times are required to strongly detect every abundance thereafter.

Due to the width of the bin narrowing as the bandwidth decreases, the amount of spectra averaged in each point lessens. Thus, when investigating two rather than three photometric points, it is no surprise that all of the results worsen, and we can confidently say that two points will not be sufficient for most observations of interest due to the lack of sufficient spectral data. However, even when looking at three photometric points, there are caveats to this idea. Again, due to the bin narrowing, increasing the resolving power yields less favorable results, with more difficult detections for varying \ce{H2O} abundances, although even still at 3\% bandwidth, a modern Earth abundance and lower is still able to be strongly detected. Narrower bandwidths are more true to the spectrum, but contain less averaged spectra per point. Additionally, when over-resolving a band, you will incur observational penalties to achieve the resolution and thus, will still not benefit from having the higher resolution. As you widen the bin, and thus lower the resolving power, \ce{H2O} becomes much more easily detectable at all abundances that are possible to strongly detect, with less observational penalty. Comparing to prior work in BARBIE1, which assumed an R of 140 (and thus, an IFS observation), a modern Earth at 0.94 {\microns} requires an normalized exposure time of 0.81 to detect - and a modern Earth at 0.94 {\microns} with bandwidth of 10\% and 3 photometric points requires an normalized exposure time of 1.22 to strongly detect. However, we cannot directly compare normalized exposure times between two different observational methods (i.e. photometry and spectroscopy); in terms of normalized exposure time, a 10\% broadband observation requires 1.22 and a spectroscopic observation at R = 140 requires 0.25. However, a photometric observation will be much less noisy and likely require less time than an IFS observation, which would likely be preferable in a first-order characterization of \ce{H2O}. We find that in every scenario, abundances lower than $\sim$$1\times10^{-3}$ VMR will likely require spectroscopic rather than photometric observations - with the feature depth so shallow, it will be incredibly difficult to constrain photometrically, and as you increase the resolving power, you grow nearer to spectroscopy observations regardless. 

Our retrieval results indicate that \ce{H2O} can be detected photometrically, at moderate normalized exposure time, if there is a minimum of 3 spectral points placed at continuum on either side of the feature depth. However, we must study the yields in conjunction with the retrieval results to understand if this is the most efficient method of observation, especially taking into account the different possible noise cases per observation. Figure~\ref{fig:yields_long} shows the huge impact that noise has on the best selection method. As expected, as the noise increases, the spectroscopic observations result in lower yields, with photometric observations resulting in higher yields. However, these calculations assumed that the wavelength for characterization was the longer wavelength point past the feature depth - this is a pessimistic assumption, as longer wavelengths increase the difficulty of observation; additionally, we have an induced noise floor which can limit the performance of photometric observations. 

Thus, we investigate other possible yield drivers in Figure~\ref{fig:yields_mult}, with the wavelength for characterization selected more optimistically for the depth of the feature, or the wavelength for characterization held at the longer wavelength point, but the noise floor reduced effectively to 0, or increasing the aperture size. We find that the wavelength for characterization, whether just past the feature or at the depth of the feature, is not a critical driver of resultant EEC yields per observational method. There are slight increases in the yields for every method and resolving power by selecting a different characterization wavelength (i.e. not having the wavelength for characterization at continuum at the long wavelength end of the band) but no large or significant changes in the resultant yields. This supports moving to shorter wavelengths within the 0.9 {\microns} would improve the overall yields and observation quality. However, the detector noise level is still a much more important driver of the yields.

Next, we investigate effectively removing the noise floor - while this does not critically impact the spectroscopic observations, the noise floor could limit the efficacy of photometric observations due to the core of the PSF, which could in turn lead to higher yields with a lower noise floor. Indeed, in Figure~\ref{fig:yields_mult}b, we see a significant increase of yields. Most interestingly, spectroscopic observations have the largest yield increases comparatively to photometric observations, when the opposite was expected. Spectroscopic observations become more preferred for higher noise cases than before, however this is not sustained as the noise instances continue to increase. As the noise case continues to increase, the 10\% bandwidth case has the highest yields among the three bandwidths considered for photometry for all noise floor levels. This is likely due to the fact that lowering the noise floor is beneficial to all observational methods, and although the photometric observations see significant improvement, spectroscopy is much more sensitive to noise. Thus, removing the noise floor is critically beneficial to spectroscopic noise observations, where the photometric observations see a diminishing return from lowering the noise floor, conversely from expected. The noise floor, however, is still not as critical a driver of yields as the detector noise level.

However, a large driver of EEC yields is the aperture size of the telescope. By increasing the aperture size as in Figure~\ref{fig:yields_mult}c, we see that all of the resultant yields leap upwards, in most cases doubling. However, the largest difference between all the presented results is the noise case at which photometry becomes preferable to spectroscopy. While in every other instance of yield calculation, $1\times10^{-3} pix^{-1} s^{-1}$ represents a flex point at which broadband, 10\% bandwidth photometric observations performs better than any spectroscopic observation and 5\% bandwidth photometric observations performs as well as or better than spectroscopic observations, this is no longer the case with an increased aperture. The noise case at which photometry is preferred increases to $1\times10^{-2} pix^{-1} s^{-1}$, an order of magnitude greater noise. At $1\times10^{-3} pix^{-1} s^{-1}$, spectroscopy at any resolving power is preferred and has higher yields than any photometric observation. At noise cases equal to or higher than $1\times10^{-2} pix^{-1} s^{-1}$, photometry has higher resultant yields, with the exception of 3\% bandwidth photometric observations which performs approximately as well as spectroscopic observations. This is completely driven by the larger aperture size. A larger aperture size results in a sharp increase in the photon collection, which greatly lessens intrinsic noise. Additionally, a larger aperture decreases the PSF size, which is critical for increasing the yields from photometric observations \citep{stark15}.

\begin{figure*}
    \centering
    \includegraphics[width=0.5\linewidth]{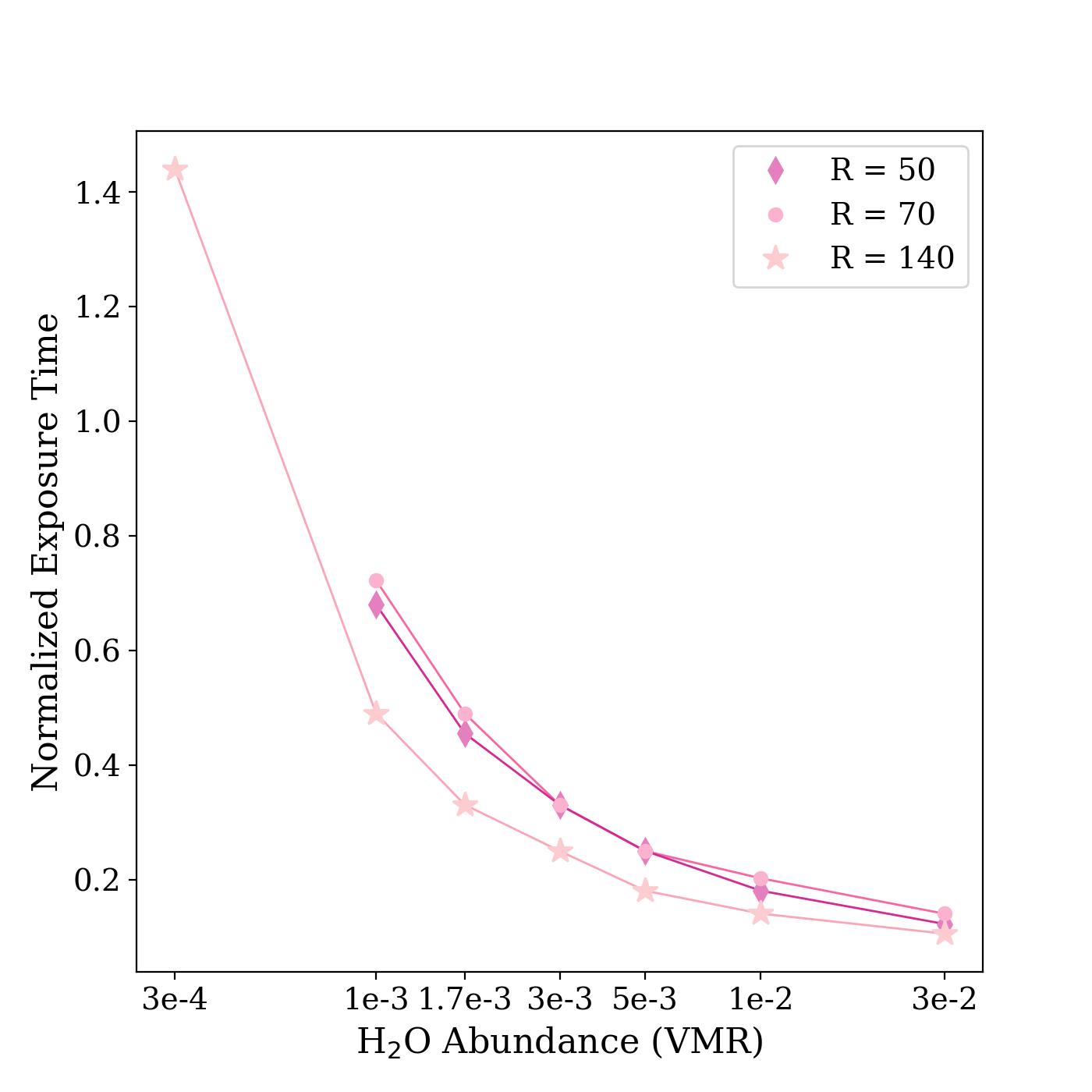}
    \caption{We present the minimum normalized exposure time required for a strong detection of \ce{H2O} at 0.9 {\microns}. The y-axis presents the normalized exposure time, the x-axis presents the varying \ce{H2O} abundances tested, and each curve represents a different tested resolving power from R = 50, 70, and 140. \textbf{At mid-level resolving powers, they will converge to nearly identical results.} }
    \label{fig:resolving_eac}
\end{figure*}

Reflected throughout this work is the relatively negligible difference in R = 50 vs R = 70 spectroscopic results. In Figure~\ref{fig:resolving_eac}, we present the minimum normalized exposure time required for strong \ce{H2O} detection for the resolving powers tested herein. We find that the difference in R for spectroscopic observations is negligible, as most mid-level resolving powers (such as 50 and 70) converge to the same required normalized exposure times. At sufficiently high abundances of \ce{H2O}, all resolving powers converge to approximately the same required normalized exposure time. Thus, observing spectroscopically with R = 50 and R = 70 not only would result in incredibly similar exposure times, but similar yields in almost all cases.

However, when applying real-world observations and expectations to these results, we are required to set the limit on a reasonable noise case. Realistically, a noise case higher than $1\times10^{-3} pix^{-1} s^{-1}$ would be far too high, and unlikely to be selected due to low performance. Lower than $1\times10^{-3} pix^{-1} s^{-1}$ is ideal and the current target - in which case, spectroscopic observations are far and away the \textbf{optimal} choice for higher yields and more information for detection and characterization of Earth-like worlds. In fact, the Roman Space Telescope EMCCD \textbf{\citep{romanemccd}} currently has a dark current (which is the only value we use in our study as a proxy for all noise components) of $3\times10^{-5} pix^{-1} s^{-1}$ - significantly lower than $1\times10^{-3} pix^{-1} s^{-1}$, and well within the realm of \textbf{top performing detectors under consideration for HWO}. $1\times10^{-3} pix^{-1} s^{-1}$ as a noise case is certainly still possible, especially when considering the possibility of space servicing to improve instruments after launch - and the best observational method here becomes less clear. The observer could select either a broadband photometric observation that would result in the highest possible yields, or could opt for a lower-resolution spectroscopic observations resulting in an $\sim$8\% decrease in yields, but higher quality data. Additionally, we note that in every case, spectroscopic observations result in the closest yields to 25 EECs, which is one of the observational goals of HWO \citep{decadal} - the only exception is the large aperture case, wherein every observational method can achieve 25 EECs depending on the noise case, however even there spectroscopic observations result in the highest yields. 

This is all presented with an important caveat, however: while the mission can fly with higher noise instruments and remain capable of detecting and characterizing \ce{H2O}, \ce{O2} detection and characterization would then no longer be possible. Since the \ce{O2} spectral feature at 0.76 {\microns} is narrow and sharp, it is significantly diluted when using a very low spectral resolution or photometric bin width, and requires relatively high-resolution spectral resolution ($\sim$140) for efficient measurement. With higher noise instances, spectroscopy yields are significantly lower, and would likely sink lower when characterizing \ce{O2} instead of \ce{H2O}. If a primary goal of HWO is to detect and characterize \ce{O2}, this drives the requirement for lower noise cases, thus again determining that spectroscopic characterization of both \ce{O2} and \ce{H2O} is preferred over photometric observations. In essence, while photometric \ce{H2O} detection is possible, and preferable under high detector noise, \ce{O2} detection requires spectroscopy. 

However, other parameters can control the relative merits of one observational method over another. The number of noise pixels per spectral resolution element would be a driver in shifting the yield curves - we assume a conservative value of 6, but there are designs that would lessen this to 2. This would shift the curves to the right, making spectroscopy more beneficial even at higher noise instances. Another factor can be the throughput penalty of an IFS - with a more efficient IFS, the throughput penalty can be reduced from 30\% to 20\% or lower, thus also driving more spectroscopic observations over photometric observations. Additionally, we assume no parallelization of observations for photometry or spectroscopy - by adding the parallel channels, the yields would change dramatically for both observational methods. We also do not assume parallel coronagraphs for the planet detection using photometry - if we assume a selectable dichroic, it would be possible to observe two spectral points at once, and then re-position the split to catch an additional spectral point. Under this assumption, the photometric yields would be more optimistic, and thus the curves would shift higher.

In future works, we will develop a more robust photometric observation simulation framework to more finely understand which molecules will benefit from being observed photometrically as opposed to spectroscopically. This will allow us to understand the foundational best practices for biosignature detection outside of spectroscopy alone, thus changing our observational procedures to include variance in characterization techniques. Additionally, we will explore different parameters and their contributions to increasing or decreasing EEC yields, including noise components, observational techniques, parallel channels, and others. More investigation is necessary to see if there are additional main drivers of yields outside of the noise case, such as those discussed, and which noise components impact the resultant yields more than others. However, it is clear that detector noise and instrument throughput play a critical role in determining the best observational method for detecting and characterizing \ce{H2O}.

\begin{figure*}
\centering
\gridline{\fig{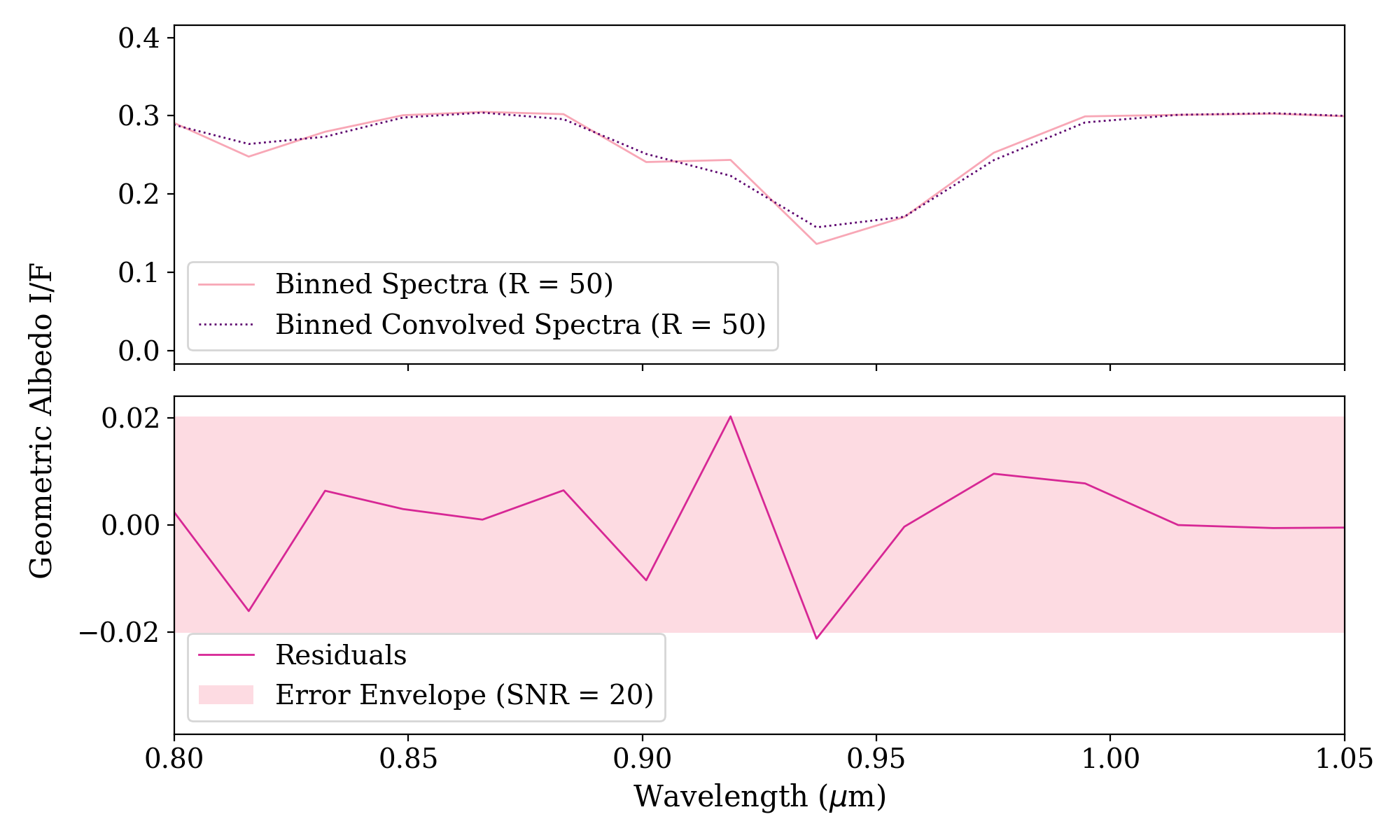}{0.48\textwidth}{(a) Spectra and residuals at R = 50.}
          \fig{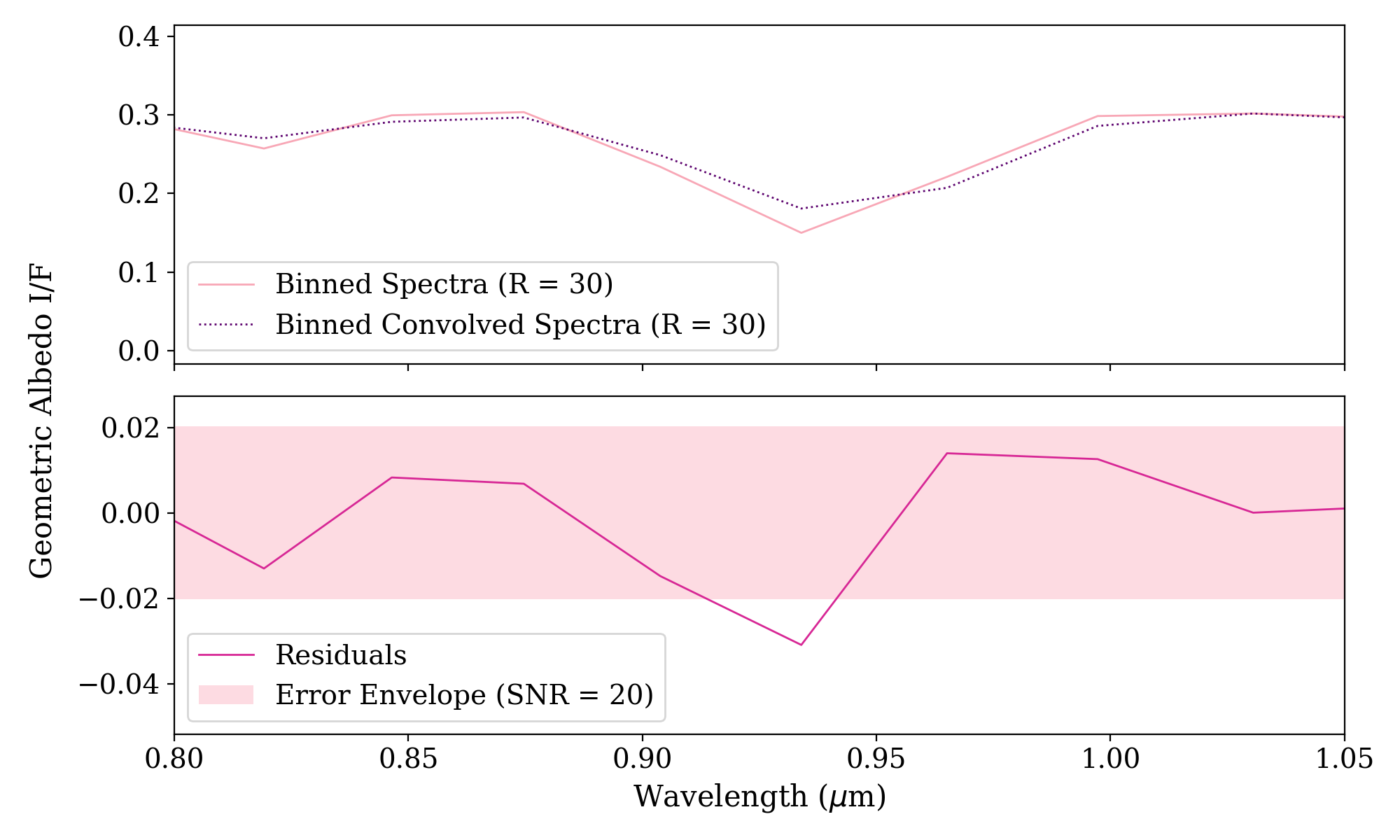}{0.48\textwidth}{(b) Spectra and residuals at R = 30.}}
\caption{In the top panel, we present two spectra: one that is binned to the desired resolving power (solid line) and one that is convolved, and then binned (dotted line). In the bottom panel, we present the residuals between the two spectra, along with the associated error envelope for an SNR of 20, the highest SNR investigated in our study. \textbf{The method of spectral binning is not a driver of error at mid-level resolving powers, but become more potent at lower resolving powers and careful selection of convolution vs binning becomes critical.}}
\label{fig:convolution}
\end{figure*}

Additionally, we note that PSG bins data from the native resolving powers to the desired resolving powers by averaging the nearby spectral points (i.e. if binning from 10,000 to 1,000, each 10 spectral points will be averaged to one point) when using the Boxcar resolution type. This will lessen the sharpness of each feature, and if the wavelength range is improperly placed, can also produce a step-function-like shape to spectral features. This will also produce an apparent improvement to the noise, which can result in a more optimistic simulation. When using a convolution approach, the feature would be broadened, lessening the native sharpness of each feature. These two methods can result in different resolving widths of the feature in question, and would require further investigation to determine the resolution of each feature and thus the required resolving power. For instance, by using PSG's Gaussian resolution type instead where the sampling of spectral points are computed at the selected resolution multiplied by 10, and then convolved with a Gaussian kernel with a full-width-half-max (FWHM) of the resolution. 

For the work contained herein, due to the width of the \ce{H2O} feature, the convolution method for the model spectra does not greatly impact the results. In Figure~\ref{fig:convolution}a, we present the spectrum of the lowest resolving power tested for spectroscopic observation (R = 50) as the binned spectrum (which was used in this work) vs the convolved spectrum that is then binned, mimicking an observation using an IFS, in the top panel. In the bottom panel, we show the residuals between the two spectra, along with the error envelope for an SNR of 20, which is the highest possible SNR in this work. We can see that the residuals do not exceed the error envelope, i.e. the difference between using a binned spectrum vs a convolved and then binned spectrum is negligible. However, there will invariably be a point in resolution space for \ce{H2O}, specifically the 0.9 {\microns} feature, wherein the difference between convolution and binning will induce additional error. To briefly investigate, we present the same analysis in Figure~\ref{fig:convolution}b, wherein all plot aspects are identical but are portraying a resolving power of 30. We can see that not only is the \ce{H2O} feature significantly less discernible by eye, the difference between binning and convolving also grows, with the residuals spilling over the SNR 20 error envelope at the presumed feature depth spectral point. We can infer that at lower and lower resolving powers, the choice to convolve or only bin will become more critical; however, at these very low resolving powers, we enter the regime of narrow band photometry, and thus convolution no longer becomes an aspect to consider. This effect will be far more pronounced for less broad features such as the \ce{O2} feature at 0.76 {\microns}; further analysis is necessary to understand the resolving power requirements of detecting and resolving the \ce{O2} feature. 

\section{Conclusions}
\label{sec:conc}

In summation, a modern Earth abundance of \ce{H2O} is strongly detectable photometrically assuming a minimum of 3 spectral points, for the bandpass bandwidths investigated (10\%, 5\%, 3\%) and moderate normalized exposure time, so long as the points are placed on either side of the continuum and one at feature depth. Indeed, abundances lower than a modern Earth (as low as $1\times10^{-3}$ VMR) are also strongly detectable with this technique. At abundances lower than $1\times10^{-3}$ VMR, spectroscopic observations are required. As more points are added, results improve, but with the caveat that the increasing resolving power becomes more conducive to a spectroscopic observation rather than a photometric observation. At higher noise cases, photometric observations result in higher EEC yields and are preferable to spectroscopic observations; at lower noise cases, the reverse is true. At the noise instance of $1\times10^{-3} pix^{-1} s^{-1}$, it appears to be a flex point where either photometric or spectroscopic observations can be preferred based on the desires of the observer and details of the technology, however a larger aperture results in better spectroscopic yields than photometric at $1\times10^{-3} pix^{-1} s^{-1}$.

N. L. gratefully acknowledges financial support by an appointment to the NASA Postdoctoral Program (NPP) at the NASA Goddard Space Flight Center, administered by Oak Ridge Associated Universities under contract with NASA. The views and conclusions contained in this document are those of the authors and should not be interpreted as representing the official policies, either expressed or implied, of the National Aeronautics and Space Administration (NASA) or the U.S. Government. The U.S.Government is authorized to reproduce and distribute reprints for Government purposes notwithstanding any copyright notation herein. The authors thank the referee for their thoughtful comments that improved the quality of this manuscript. N. L. also gratefully acknowledges Greta Gerwig, Margot Robbie, Ryan Gosling, Emma Mackey, and Mattel Inc.{\texttrademark} for Barbie (doll, movie, and concept), for which the BARBIE project is named after. This Barbie is an astrophysicist! The authors would like to thank Dr. Susan Redmond for her Coronagraph design used in the AYO yield calculations. The authors would also like to thank the Sellers Exoplanet Environments Collaboration (SEEC) and ExoSpec teams at NASA's Goddard Space Flight Center for their consistent support.

\bibliographystyle{aasjournal}
\bibliography{main}

\end{document}